\documentclass{jfm}
\usepackage{graphicx}
\usepackage{epstopdf, epsfig}
\usepackage{amsmath}
\usepackage{color}
\usepackage{rotating}

\begin{document}

\newtheorem{lemma}{Lemma}
\newtheorem{corollary}{Corollary}

\shorttitle{Inertial migration of spherical and oblate particles in straight ducts} 
\shortauthor{I. Lashgari, M.N. Ardekani, I.Banerjee, A.Russom and L. Brandt} 

\title{
Inertial migration of spherical and oblate particles in straight ducts}

\author
 {
 Iman Lashgari\aff{1}
  \corresp{\email{imanl@mech.kth.se}},
 Mehdi Niazi Ardekani \aff{1},
 Indradumna Banerjee \aff{2},
 Aman Russom \aff{2}
  \and 
  Luca Brandt\aff{1}
  }

\affiliation
{
\aff{1}
 Linn\'e FLOW Centre and SeRC (Swedish e-Science Research Centre),KTH Mechanics, SE-100 44 Stockholm, Sweden
\aff{2}
Division of Proteomics and Nanobiotechnology, KTH Royal Institute of Technology, Stockholm, Sweden
}

\maketitle

\begin{abstract}
We study numerically the inertial migration of a single rigid sphere and an oblate spheroid in straight square and rectangular ducts. A highly accurate interface-resolved numerical algorithm is employed to analyse the entire migration dynamics of the oblate particle and compare it with that of the sphere. 
Similarly to the inertial focusing of spheres, the oblate particle reaches one of the four face-centred equilibrium positions, however they are vertically aligned with the axis of symmetry in the spanwise direction. In addition, the lateral trajectories of spheres and oblates collapse into an equilibrium manifold before ending at the equilibrium positions, with the equilibrium manifold tangential to lines of constant background shear for both sphere and oblate particles. The differences between the migration of the oblate and sphere are also presented, in particular the oblate may focus on the diagonal symmetry line of the duct cross-section, close to one of the corners, if its diameter is larger than a certain threshold. Moreover, we show that the final orientation and rotation of the oblate exhibit a chaotic behaviour for Reynolds numbers beyond a critical value. 
{Finally, we document that the lateral motion of the oblate particle is less uniform than that of the spherical particle due to its evident tumbling motion throughout the migration. In a square duct, the strong tumbling motion of the oblate in the first stage of the migration results in a lower lateral velocity and consequently longer focusing length with respect to that of the spherical particle. The opposite is true in a rectangular duct where the higher lateral velocity of the oblate in the second stage of the migration, with negligible tumbling, gives rise to shorter focusing lengths.} 
These results can help the design of microfluidic systems for bio-applications.

\end{abstract}

\section{Introduction}
Inertial focusing is an effective technique in micro-fluidics to control the motion of micron size particles transported by a fluid \cite[][]{Toner14}. One advantage of this approach is that the particle motion is governed by intrinsic hydrodynamic forces while non-hydrodynamic forces such as magnetic, optic and acoustic forces are often absent \cite[passive microfluidic as denoted by][]{Hur11}. From the practical point of view, inertial focusing is therefore rather simple to use, i.e. no additional component is needed, with the potential of high throughput. The deterministic ordering induced by inertial particle focusing can be employed directly in many biological applications such as large-scale filtration systems, continuous ordering, cell separation and high throughput cytometry \cite[][]{Bhagat08,Zhou13a}. In this study we employ an interface-resolved numerical algorithm based on the Immersed Boundary Method to provide deeper and more accurate understanding of the physics of the particle motion in micro-devices. For the first time, the entire migration dynamics of an oblate particle is analysed and compared to that of a sphere. This knowledge can be directly used to develop methods for highly efficient inertial focusing, with high purity and throughput, without compromising the operational simplicity.%

In the absence of inertia, particle trajectories follow the flow streamlines to preserve the reversibility of the Stokes flow. When inertia becomes relevant, however,
 particle lateral migration is observed and is directly connected to the non-linear dynamics of the underlying fluid flow. In inertial flows, the ratio between the inertial to viscous force is measured by the Reynolds number, $Re=UH/\nu$, where U and H are the characteristics velocity and length scales of the flow and $\nu$ is the fluid dynamic viscosity. In typical micro-fluidic applications, when $Re > 1$,  inertial forces govern the lateral motion and final arrangement of the particles. 

Particle lateral motion and focusing were probably first observed in the pipe experiment by \cite{Segre61}. Later on, few theoretical studies aimed to explain the particle migration: among others \cite{Rubinow61} propose a formulation for the lateral force based on the particle rotation while \cite{Saffman65} suggests a lift force as function of the particle relative velocity with respect to the undisturbed fluid velocity at the particle position. The latter is shown to be independent of the particle rotation. During the last 20 years and owing to the  increase of inertial micro-fluidics  for biomedical applications, the number of theoretical, numerical and experimental investigations on inertial particle migration has significantly increased. The slip-spin force of  \cite{Rubinow61} and the slip-shear force of \cite{Saffman65} are denoted as weak forces in the work by \cite{Matas04IFP}. The dominating forces, determining the particle lateral trajectory and the equilibrium position,  are found to be shear-induced and wall-induced lift forces resulting from the resistance of the rigid particle to deformation \cite[see the review by][]{Toner14}. The wall-directed shear-induced lift force and the centre-directed wall-induced lift force balance each other at the equilibrium position observed in the experiment by \cite{Segre61}. 

Theoretical analyses of the lateral force on a particle in simple shear and channel flow have also been conducted after the seminal work by \cite{Segre61}. Among others, \cite{Ho74} predicts the trajectory and equilibrium position of a rigid spherical particle in a two-dimensional Poiseuille flow using analytical calculations that compared well with the experiment. Particle lateral motion is studied extensively in the framework of matched asymptotic theory as introduced by \cite{Schonberg89} and extended by \cite{Asmolov99} to account for the flow inertia. This theory is based on the solution of the equations governing the disturbance flow around a particle. Following the work by \cite{Asmolov99}, many analytical, numerical and experimental studies have tried to determine the lateral force on the particle including Reynolds number and particle size effects. As examples \cite{Matas04} and \cite{Matas09} report the lift force on the particle across the pipe as a function of the Reynolds number. Similar to the theoretical predictions, they observe that particles form a narrow annulus which moves toward the wall as Re increases. However, they find additional equilibrium positions closer to the pipe centre which are not predicted by theory and are attributed to the finite-size effect. Finally, they note that experimental results deviate from the theoretical prediction as the ratio between the channel height and the particle diameter becomes smaller.

The majority of the studies related to microfluidics applications are conducted in duct flow due to the simplicity of microfabrication methods  with material such as PDMS. In addition, the geometrical anisotropy (radial-asymmetry) of the duct  is the key to isolate and focus the particles in specific  positions over the cross section. 
It is well known that particles in square duct migrate to four face-centred equilibrium positions for small and moderate values of the Reynolds number \cite[see among others][]{Chun06,Dicarlo07}. New equilibrium positions are also reported mainly when the Reynolds number is beyond a certain threshold \cite[see][]{Chun06,Bhagat08,Miura14}. The lateral forces on a particle in a duct changes in magnitude and direction across the cross-section and also depend on  the particle size and bulk Reynolds number. Therefore, the physical mechanisms behind the particle lateral motion in ducts is not yet fully understood as recently concluded in the experimental work by \cite{Miura14}.

Theoretical approaches aiming to predict the particle lateral motion in a duct are challenging due to the three-dimensionality of the driving flow. 
\cite{Dicarlo09} report that the lift force on a particle moving in a duct can collapse into a single curve in the near wall region or close to the duct centre depending on the choice of scaling. In particular, they multiply the drag coefficient by the ratio between particle diameter to duct height, $D/H$, or divide it by $(D/H)^2$ to get the collapsed data near wall or close to the centre. Recently \cite{Hood15} have extended the theoretical approach of \cite{Ho74} to express the lateral force on a particle in a three-dimensional Poiseuille flow as a function of the particle size. Including terms due to the wall contribution they show how a larger particle focuses closer to the centre, in agreement with their numerical experiments. These theoretical studies are however limited to spherical particles because there is no asymptotic analytical solution for non-spherical particles and in most of the cases it is assumed that the particle does not affect the underling flow. In addition, it is challenging to measure in an experiement the entire dynamics of the particle motion, such as trajectory, velocity, orientation and rotation rate. Therefore, interface-resolved numerical simulations, where the full interactions between the fluid and solid phase are taken into account, have become a valuable way to study inertial migration of particles of different shape. Numerical simulations may also provide information about the role of different control parameters such as particle size, shape and Reynolds number on the particle migration.

In this study we simulate the motion of rigid particles in straight ducts of different aspect ratios employing an Immersed Boundary Method. To the best of our knowledge this is the first study on the motion of both spherical and oblate particles in a microfluidic configuration where not only the equilibrium position but also the entire migration dynamics of the particle from the initial to final position,  including particle trajectory, velocity, rotation and orientation, are investigated. The paper is organised as follows. In \S 2 we introduce the governing equation, numerical method and flow setup and in \S 3 discuss the results of simulations of spheres and oblate spheroids in a square and rectangular duct. The main conclusions are summarised in \S 4. 

\section{Methodology}
In this section we discuss the governing equations, numerical method and the flow setup for the simulation of a single rigid particle in a straight duct. 
\subsection{Governing equations}
\begin{figure}
\begin{center}
   \includegraphics[width=6cm]{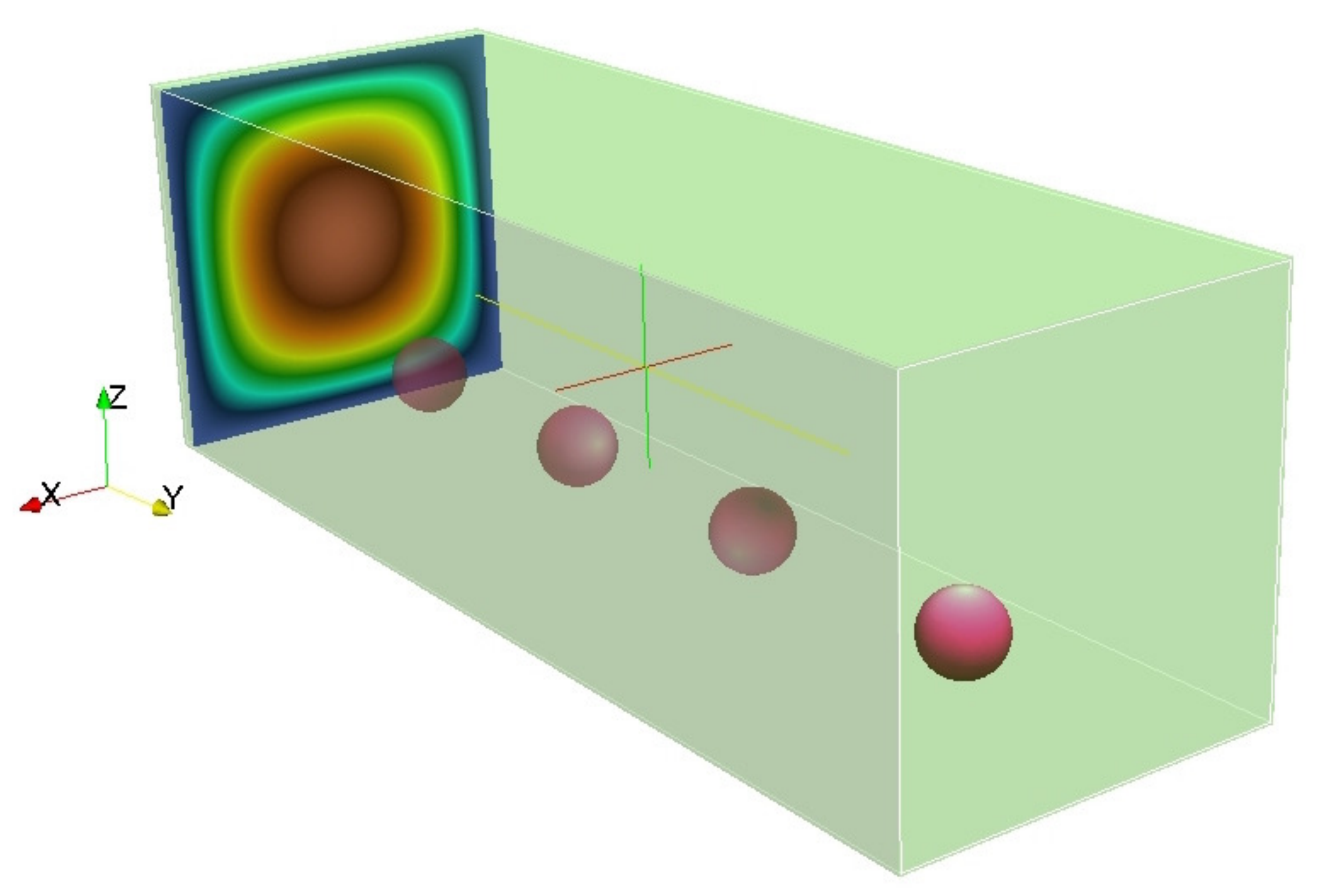}
   \includegraphics[width=6cm]{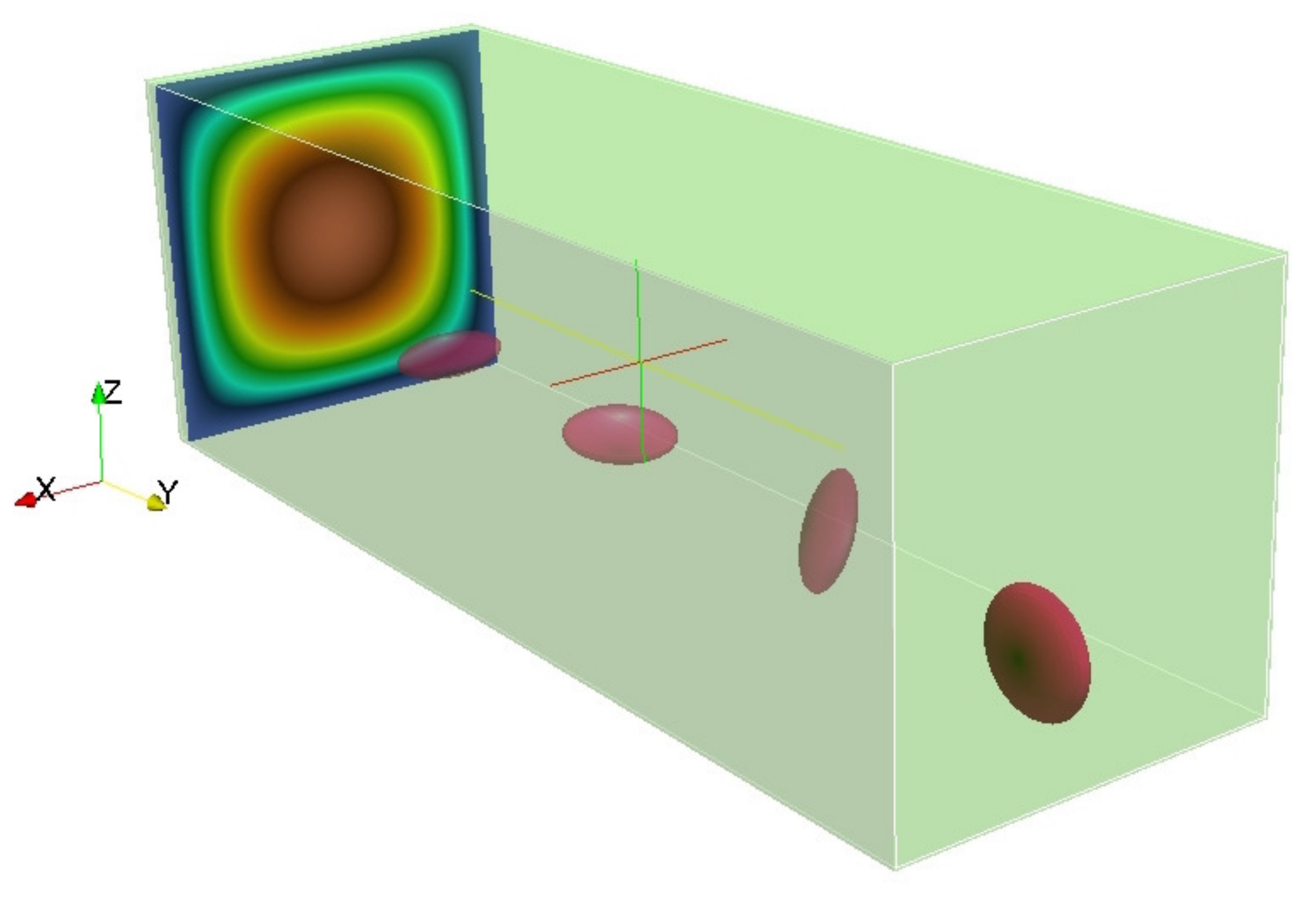}
   \caption{a) Flow visualisation of the motion of a single spherical/oblate particle in a square duct. The particle and duct size are shown at the actual scale. }
   \label{fig:fig1}
\end{center}
\end{figure}

Inertial flows, unlike Stokes flow, are described by a nonlinear governing equation such that the flow system is irreversible. Therefore, the trajectory of a particle does not necessarily follow the streamlines and the particle experiences lateral motion. In this study the fluid is incompressible and Newtonian and its motion is governed by the Navier-Stokes and continuity equations,
\begin{align}
\rho (\frac{\partial \textbf{u}}{\partial t} + \textbf{u} \cdot \nabla \textbf{u}) = -\nabla P + \mu \nabla^2 \textbf{u} + \rho \textbf{f}, \\ \nonumber
\nabla \cdot \textbf{u} = 0,
\label{eq:NS}  
\end{align}
where $P$ and $\mu$  indicate the fluid pressure and dynamic viscosity and $\rho$ the fluid density. Since we simulate the motion of a neutrally buoyant particle, $\rho$ denotes also the particle density. We use the coordinate system and velocity component $\textbf{X}=(x,y,z)$ and $\textbf{u}=(u,v,w)$ corresponding to the streamwise and cross flow directions as shown in figure~\ref{fig:fig1}. A force $\textbf{f}$ is added on the right hand side of the Navier-Stokes equation to model the presence of the finite size particle. The motion of the rigid particle is governed by the Newton-Euler equations,
\begin{align}
m^p \frac{ d \textbf{U}_c^{p}}{dt} = \oint_{\partial {V}_p}  [ -PI + \mu (\nabla \textbf{u} + \nabla \textbf{u}^T ) ] \cdot  \textbf{n} dS+ \textbf{F}_c , \\ \nonumber
I^p \frac{ d \pmb{\Omega}_c^{p}}{dt} = \oint_{\partial {V}_p}  \textbf{r} \times \big{\{} [ -PI + \mu (\nabla \textbf{u} + \nabla \textbf{u}^T ) ] \cdot  \textbf{n}  \big{\}} dS + \textbf{T}_c, 
\label{eq:NE}  
\end{align}
where $m^p$ and $I^p$, $\textbf{U}_c^p$ and $\pmb{\Omega}_c^p$ are the mass, moment inertia, centre velocity and rotation rate of the particle $p$. The surface of the particles and unit normal vector are denoted by $\partial {V}_p$ and  $\textbf{n}$ whereas the vector connecting the centre to the surface of the particles by $\textbf{r}$. The first term on right hand sides of these equations represents the net force/moment on particle $p$ resulting from hydrodynamic interactions.  The second term, $\textbf{F}_c$ and $\textbf{T}_c$, indicates the force and torque resulting from contact interactions. Note, however, that during the entire motion of a particle in the duct from initial to final position, we do not observe any particle-wall contact for the cases presented here. 
An interface condition is needed to enforce the fluid velocity at each point of the particle surface to be equal to the particle velocity at that point,  $\textbf{u}(\textbf{X})= \textbf{U}^p (\textbf{X}) = \textbf{U}_c^p + \pmb{\Omega}_c^p \times \textbf{r}$. An Immersed Boundary Method with direct forcing is employed to satisfy the interface condition indirectly by applying the forcing  $\textbf{f}$ in the vicinity of the particle surface.

\subsection{Numerical method}
The numerical approach is a combination of a flow solver with an Immersed Boundary Method.  The fluid flow is computed by discretising the governing equations on a staggered mesh using a second-order finite different scheme. The fluid/solid interaction is based on the discrete forcing method, a variant of the IBM \citep{Mittal05}, developed by \cite{Uhlmann05} to simulate finite size particle suspensions. This has been further modified by \cite{Breugem12} to simulate neutrally buoyant particles with second-order spatial accuracy. Two sets of grid points are considered: an equi-spaced fixed Eulerian mesh everywhere in the domain and Lagrangian grid points uniformly distributed on the surface of the particle. The Eulerian and Lagrangian grid points communicate to compute the IB forcing which ensures the no-slip and no-penetration boundary conditions on the particle surface. The IB force is applied on both fluid and particle phase when the velocities and positions are updated.  We note for the sake of completeness that even though the near-field interactions between the two particles or particle-wall are irrelevant in this work because we simulate the motion of a single particle that is always located far from the walls, the numerical algorithm contains both lubrication correction and soft sphere collision model for small gaps between the solid objects \cite[see the Appendix of][for more details]{Lambert13}. The accuracy of the code has been examined in \cite{Breugem12,Costa15} where good agreement with experimental data is obtained. In addition, the present implementation has been employed over a wide range of Reynolds numbers and particle concentrations in different studies on rigid particle suspension for both spherical and non-spherical particles \cite[see for example][]{Lashgari14,Picano15,Fornari16a,Niazi16a,Niazi16b}.

\subsection{Flow setup}
In this study the motion of a rigid sphere and an oblate spheroid are examined in square and rectangle cross-section straight ducts. The no-slip and no-penetration boundary conditions are imposed on four lateral walls with periodic boundary condition in the streamwise direction. The square duct, of height $H$,  is 3H long and 
is meshed by $480 \times 160 \times 160$ Eulerian grid points in the streamwise and cross-flow directions. The rectangular duct has aspect ratio 2 and is obtained doubling the length of two parallel walls. The number of grid points is therefore  $480 \times 320 \times 160$. The number of Eulerian grid points per particle diameter is $32$ whereas $3219$ and $3720$ Lagrangian grid points are spread over the surface of the spherical and oblate particles to ensure that the interactions between the fluid and particle are fully captured. We perform the simulations with constant mass flux through the duct. The majority of the simulations are performed at $Re_b=HU_b/\nu=100$ where H is the duct height (the same for the square and rectangular duct), $U_b$ is the bulk velocity and $\nu$ is the kinematic viscosity of the fluid. This value of the Reynolds number is relevant in practical applications considering the dimensions and flow rate in micro-channels \cite[][]{Amini14}. Note that we first simulate the flow in square and rectangular ducts without particle to get the fully-developed laminar solution, then introduce a single particle at different initial positions and monitor its motion through the duct.
{We also run a simulation where both fluid and particle velocities are started from zero and observe that except at the very beginning of the particle motion the entire dynamics of the particle inertial migration remain unchanged.} 
In figure \ref{fig:fig1} we display
 the evolution of  a single rigid sphere and one oblate particle in a square duct where the particle is shown at the actual size. 
 The particle experiences rotation and lateral displacement while traveling downstream through the duct and finally focuses on a particular lateral position. 
 The specification of the particles used in the different simulations will be presented when discussing the corresponding results. {Note finally that we perform box-size and resolution studies to ensure that our results are not affected by any numerical artefact, see Appendix A. In the Appendix A, we also report a validation against the data presented in \cite{Dicarlo09}}. Considering the high resolution adopted and the relatively small particle lateral velocity, one simulation may need up to two weeks employing between $32$ and  $48$ computational cores, depending on the value of the focusing length needed to reach the final steady state.   

\section{Results}
In this work we examine the lateral motion of rigid particles in both square and rectangular straight ducts employing the numerical algorithm just introduced. We compare the entire migration dynamics, particle trajectory and final equilibrium position of sphere and oblate particles to shed more light onto the mechanisms of particle lateral displacement and provide useful knowledge for the design of micro-fluidic systems. In the first step, we focus on the motion of a spherical rigid particle in square and rectangular ducts and compare our results with previous findings. For the most part of the manuscript, we study the inertial migration of a single oblate particle. When doing so, we also investigate Reynolds number and size effects on the equilibrium position and focusing length of both types of the particles.  

\subsection{Migration dynamics of a spherical particle}
There have been several experimental and numerical studies on the inertial migration of a spherical particle in duct in the last years. However, few studies are devoted to the entire migration dynamics, particle trajectories in the duct cross-section, focusing length, Reynolds number and size effects. In this section we discuss the overall behaviour of an inertial spherical particle in a duct and compare it with that reported in previous works.  

\subsubsection{Sphere in square duct} \label{SSC}
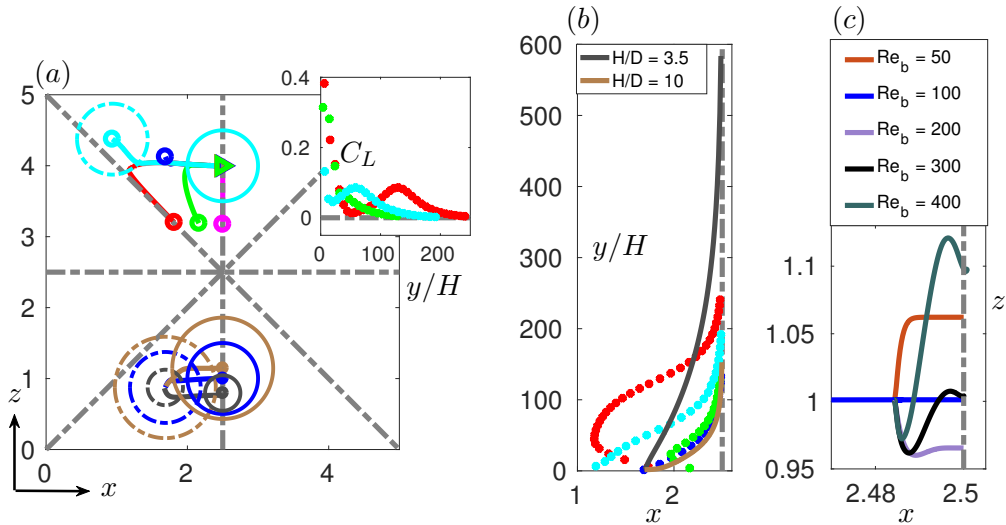
\begin{figure}
\begin{center}
   \includegraphics[width=13cm]{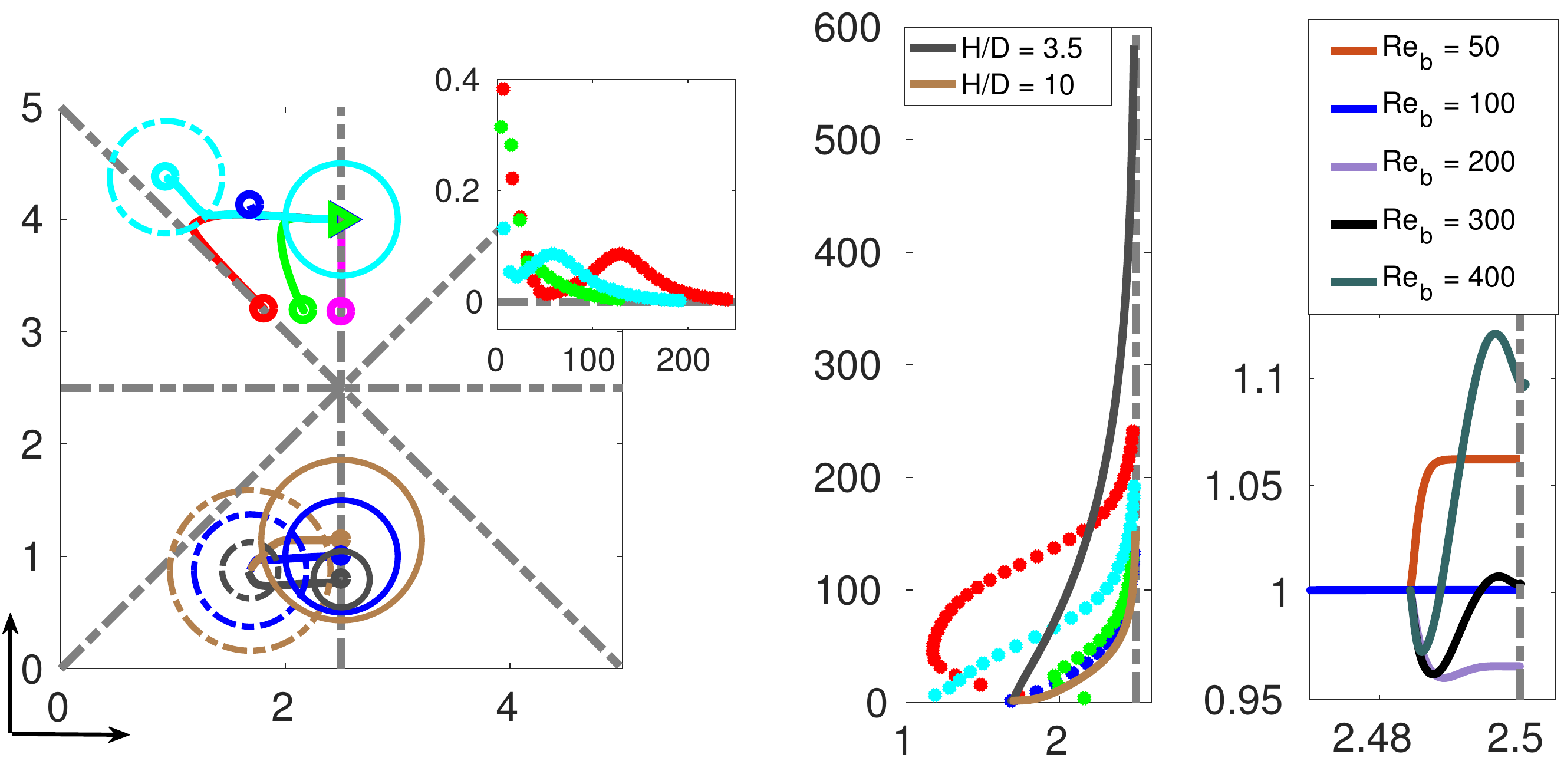}
   \put(-360,165){{\large $(a)$}} 
        \put(-335,10){{\large $x$}} 
        \put(-370,45){{\large $z$}} 
         \put(-245,135){{\large $C_L$}} 
        \put(-220,85){{\large $y/H$}}  
   \put(-160,185){{\large $(b)$}} 
         \put(-150,100){{\large $y/H$}}  
         \put(-130,-0){{\large $x$}}  
    \put(-60,185){{\large $(c)$}} 
      \put(-35,-0){{\large $x$}} 
      \put(-0,80){{\large $z$}} 
   \caption{a) Upper side of duct cross-section: Lateral trajectory of a spherical rigid particle in $1/8^{th}$ of a square duct at $Re_b=100$. The ratio between the duct height to particle diameter is $H/D_s=5$. Open circles and triangles show the starting and ending points of the lateral trajectories. We show as an example the initial and final position of a sphere with light blue dashed and solid lines. Bottom: Particle size effect on the lateral trajectory and the equilibrium position of a single sphere in square duct at $Re_b=100$. The dashed and solid lines display the initial and final position of the spheres. Right side of the cross-section: {Lift coefficient versus downstream traveling length for some of the particle trajectories.} 
 b) Focusing length for different particle trajectories (symbols are displayed at a fixed time interval). The focusing length of larger and smaller spheres is also shown by solid lines. c) Reynolds number dependence of the equilibrium position of a spherical particle with size ratio $H/D_s=5$.}
   \label{fig:fig2}
\end{center}
\end{figure}

We start with the simulation of a single spherical rigid particle in a square duct. Most of the simulations are performed for a size ratio, the ratio between the duct height (width) to the sphere diameter, $H/D_s=5$. We define the average particle Reynolds number by $\overline{Re_p}=\overline{\dot \gamma} D_s^2/\nu$ where $\overline{\dot\gamma}={2U_b}/{D_h}$ is the average shear rate in the duct and $D_h=2WH/(W+H)$ is the hydrodynamic diameter of a duct with height, $H$, and width, $W$. Note that $D_h=H$ for a square duct. With these definitions, the bulk Reynolds number, $Re_b=100$, corresponds to the average particle Reynolds number, $\overline{Re_p}=2 Re_b(D_s/H)^2 = 8$ when $H/D_s=5$. Due to the geometrical symmetry, we vary the initial position of the particle only over $1/8^{th}$ of the duct cross section. In figure \ref{fig:fig1}(a), top panel, we show the sphere lateral trajectory for different simulations where open circles and triangles represent the start and end positions of the particle centre, respectively. We show as an example the initial and final position of the particle projection onto the cross-section using dashed and solid light blue lines. We clearly observe that all the trajectories end close to the centre of a duct wall, a face-centred equilibrium position, independent of the initial position of the particles. This position is one of the four equilibrium positions observed in the square duct already in previous works, among others \cite{Chun06} and \cite{Dicarlo07}. For $Re_b=100$ and size ratio $H/D_s=5$ we report the equilibrium position to be about $0.2H$ away from the wall of the duct, i.e. $0.3H$ from the centre.

Owing to the absence of radial symmetry in ducts, the lateral motion of the particles cannot be simply explained by the balance between shear-induced and wall-induced lift forces. In duct flows, the particle lateral motion, from the initial to the final equilibrium position, can be seen as a two-step process: 1) The fast migration towards the equilibrium wall, the wall adjacent to the equilibrium position, resulting from the interplay between the strong shear-induced and wall-induced lift forces. 2) The slow migration towards the equilibrium position induced by the weak rotation-induced lift force  \cite[see also][]{Choi11,Zhou13a}. Interestingly, the trajectories collapse into a single curve, denoted as equilibrium manifold, during the second slow lateral motion towards the equilibrium position. The equilibrium manifold is a unique lateral path leading to the equilibrium position regardless of their initial positions. Even though the existence of this manifold has been reported in the recent works by \cite{Prohm14} and \cite{Liu15}, the underlying physical mechanism is still not explained. We will attempt an explanation in connection with the results of the particle migration in a rectangular duct.

We display the effect of the size ratio, $H/D_s$, on the equilibrium position of a sphere in a square duct in the lower part of figure \ref{fig:fig2}(a). We simulate the motion of the spheres of three different sizes at $Re_b=100$ starting from the same lateral position in the duct cross section and observe a similar behaviour: the particles focus at closest face-cantered equilibrium position. At equilibrium, the larger particle are closer to the center. This is mainly attributed to the steric effect as discussed by \cite{Amini14}.  The final location of spheres with size ratio $H/D_s=[3.5,5,10]$ is $[0.27H, 0.3H, 0.34H]$ from the duct centre. The distribution  of different size at the equilibrium position resembles the schematic diagram of \cite{Hood15} based on the theoretical prediction of the lateral force on the particles.  

To better understand the origins of these lateral motions,  we examine the particle lift coefficient in the duct. Following the formulation {introduced by \cite{Asmolov99}}, we express the lift coefficient of a spherical particle as
\begin{equation}
C_L= \frac{F_L}{\rho \overline{\dot\gamma}^2 a^4},
   \label{eq:CL}
\end{equation}
where $F_L$ is the lateral force on the particle {in the opposite direction of its cross flow velocity}, $\rho$ is the density of the fluid and particle and $a$ denotes the particle radius. Since the particle lateral velocity, $U_L$, is low the force on the particle can be estimated by Stokes law, $F_L=6\pi \mu a U_L$, where $\nu$ is the fluid dynamic viscosity {\cite[][]{Zhou13a}}. Combing with equation (\ref{eq:CL}), the lift coefficient can be written as
\begin{equation}
C_L= \frac{6\pi \mu H^2}{4 \rho U_b ^2 a^3} U_L = \frac{3\pi}{2Re_b} (\frac{H}{a})^3 \frac{U_L}{U_b}.
\end{equation} 
{This is estimated employing the particle lateral velocity from the simulation data for a particle moving through the duct. The lift coefficients, corresponding to some particle trajectories, are shown in figure \ref{fig:fig2} versus the downstream distance, right part.
The profiles reflect the initial acceleration and final deceleration of the particle toward the equilibrium position due to the shear-induced and wall-induced lift forces. In particular, the particle experiences the highest lift coefficient at the beginning of its lateral motion, especially if it is started close to a diagonal symmetry line; i.e. red trajectory.}

In figure \ref{fig:fig2}(b) we show the downstream length needed for a sphere to focus, focusing length, as a function of the distance from the vertical duct symmetry line. For each particle trajectory the same colour as in panel (a) is used. As regards spheres with size ratio $H/D_s=5$, we observe that if the particle is initially located close to the duct centre and in particular close to the diagonal symmetry line, the focusing length is longer. For the particular particle trajectory indicated by the red solid line, the focusing length is about 250 times the channel height. Considering the particle motion on the longest possible trajectory towards the equilibrium position, we can predict the minimum  streamwise length required for the particle focusing. In addition, we note that if the particle starts close to the duct centre its lateral motion shows the classical slow and fast motion (the symbols represent the same time intervals). The opposite is true if the particle starts close to the wall; in this case, the lateral motion is more uniform. Considering spheres of different sizes, we report that small spheres require longer ducts, about 600 times the channel height, to focus for the same bulk Reynolds number. This behaviour is exploited in the design of a microfluidic system to efficiently separate small and large particles \cite[][]{Zhou13b}.

Next we examine the effect of Reynolds number on the equilibrium position of a sphere in a square duct. {It is known that  the particle in the centre of the channel always tends to move toward the wall regardless of Reynolds number. The final equilibrium position however gets closer to the wall with increasing the Reynolds number. This effect is first observed by \cite{Segre62} and then studied theoretically by \cite{Schonberg89} for $Re_b < 75$ and by \cite{Asmolov99} for larger $Re_b$. This effect is attributed to the narrowing of the wake behaind the particle so that the wake no longer feels the wall as the particle Reynolds number increases.}
As a consequence, the interaction between the wall and the particle decreases and the particle migrates towards the wall \cite[see also][]{Liu15}. In order to examine the Reynolds number effects, we continue a simulation where the particle is close to the equilibrium position and vary the bulk Reynolds number. We show  the final part of the particle trajectories {(z-component versus x-component)} as a function of the Reynolds number in figure \ref{fig:fig2}(c). Interestingly, we observe a non-monotonic behaviour of the final equilibrium position of the particle. As we increase the bulk Reynolds number up to 200, the particle ultimately focuses closer to the wall however at $Re_b \ge 300$ the trend reverses and the particle focuses closer to the centre. Note that we have also considered two additional cases at $Re_b=200$ and $Re_b=300$ varying the initial position of the sphere and obtain the same result. {The non-monotonic positioning of the particle as a function of Reynolds number is also observed in the experiment by \cite{Ciftlik13} in a rectangular duct. These authors also report the threshold value of $Re_b\approx 300$ for the reversing of the behaviour}. This observation is however opposite to the results of the numerical study by \cite{Chun06} and \cite{Liu15} where the particle equilibrium position moves monotonically toward the wall as Reynolds number increase. {The origin of this discrepancy is still not clear and requires further investigations}. The final location of sphere at $Re_b=[50,100, 200, 300, 400]$ are $[0.287H, 0.3H, 0.307H, 0.3H, 0.281H]$ from the channel centre. Finally we note that the lateral force on the particle increases with the Reynolds number, i.e.  shorter focusing length is needed.  
{As example the particle focusing length, when staring from the initial position shown by dark-blue line in figure \ref{fig:fig2}(a), is [150,87,55] in units of $H$ at $Re_b=[100, 200, 300]$.}

\subsubsection{Sphere in rectangular duct}

\begin{figure}
\begin{center}
   \includegraphics[width=12cm]{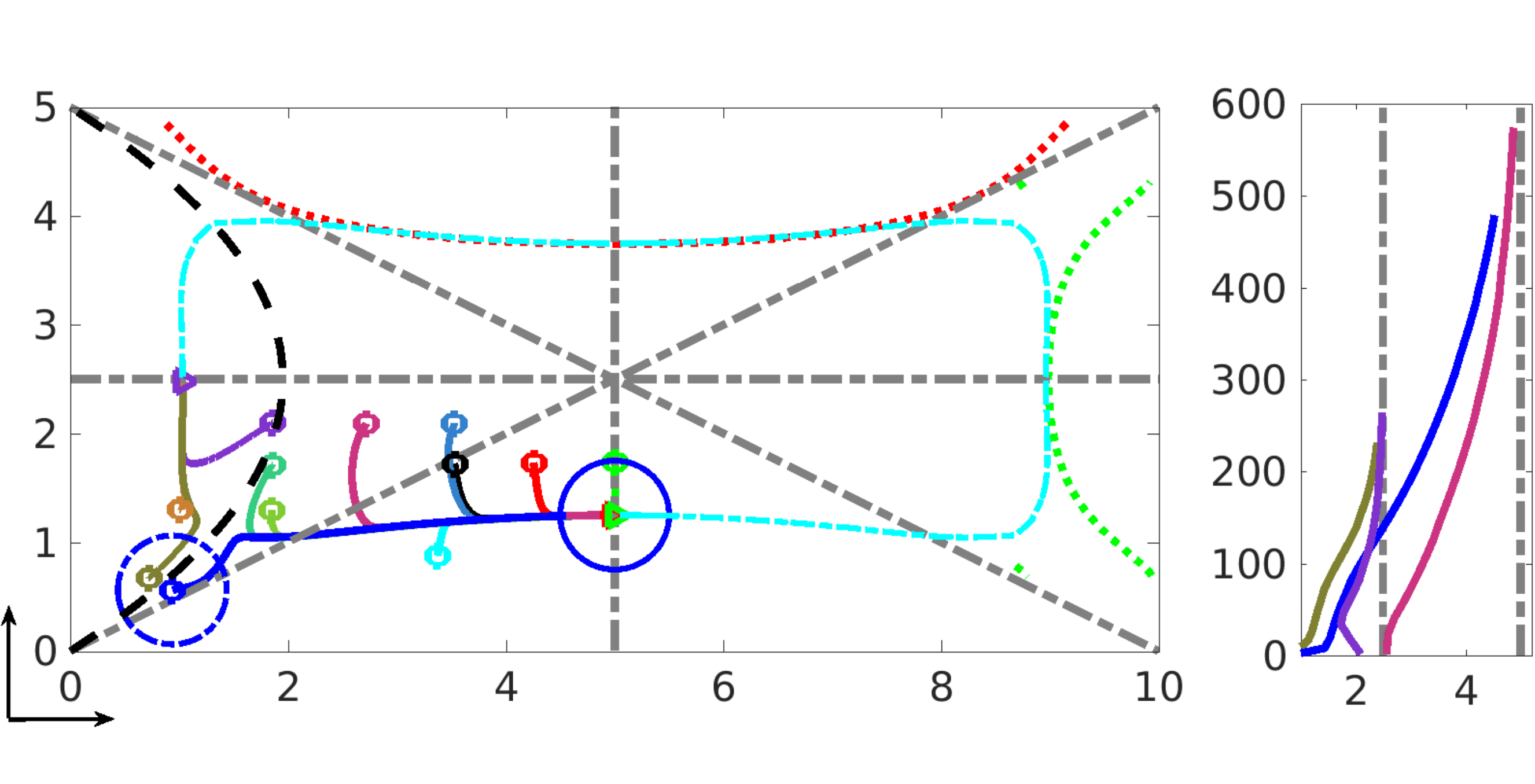}
    \put(-310,11){{\large $x$}} 
    \put(-345,45){{\large $z$}} 
   \put(-330,165){{\large $(a)$}} 
    \put(-70,165){{\large $(b)$}} 
  \put(5,90){{\large $y/H$}} 
   \caption{a) Lateral trajectories of a spherical rigid particle over $1/4^{th}$ of a rectangular duct at $Re_b=100$. The particle diameter is $1/5$ of the duct height whereas open circles and triangles show the initial and final equilibrium position of different realisations. The light-blue dashed line indicates the equilibrium manifold over the entire cross section. The black dashed line defines the initial position leading to two different equilibrium positions, at the face-centre of the long or of short wall as indicated by teh individual trajectories
   (only shown on one side).The dotted red and green lines indicate iso-levels of total shear rate equal to $0.75$ and $0.81$ and are tangential to the equilibrium manifold (only shown on one side).  {b) Focusing length for different particle trajectories. The dashed lines represent the horizontal (z = 2.5) and vertical (x = 5) duct symmetry line, as the particle focuses on either of them.}}
   \label{fig:fig3}
\end{center}
\end{figure}
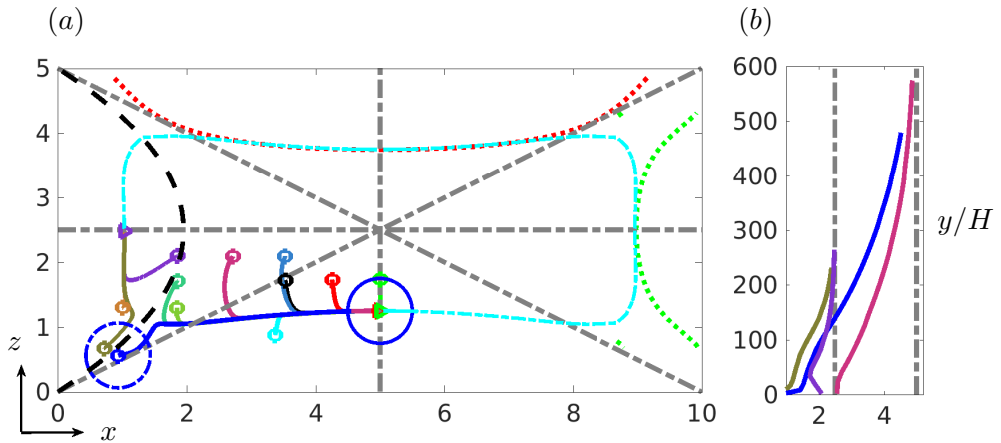
%

We next study the lateral motion of a single spherical particle in a duct with aspect ratio $2$. 
The objective is to examine the effect of the shear and lift force asymmetry on the inertial migration of the particles. 
The results are summarised in figure \ref{fig:fig3}(a) where we show the initial position, trajectory and final position of particles from different realisations exploiting the symmetry over $1/4^{th}$ of the cross section. 
The size ratio between the duct height and particle diameter is $H/D_s=5$, indicated by the blue circles in the figure, and the bulk Reynolds number $Re_b=100$ as for the results in  the previous section. 
Since the hydrodynamic diameter of the rectangular duct $D_h=4H/3$, the average particle Reynolds number $\overline{Re_p}=\overline{\dot \gamma} D_s^2/\nu=6$ for these simulations. 
In the rectangular duct, the shear-induced lift force is stronger towards the longer wall. 
Therefore, the general trend is that the particle first tends to move laterally toward the long wall and then slowly along the long wall towards the equilibrium position at the face-centre \cite[A similar behaviour is observed among others in ][]{Zhou13b,Ciftlik13, Liu15}. The particle may also focus along the symmetry line at the face-center of the short wall if it is initially located close enough to that point  \cite[][]{Bhagat08}. The final distance between the sphere centre and the centre of the rectangular duct is about 0.275H and 0.82H at the equilibrium position close to the longer and shorter wall respectively. 
The boundary delimiting the initial positions leading to the two different equilibrium positions, depicted by the dashed black line in the figure, is extrapolated from the data by fitting a parabola.

Similar to the behavior in the square duct, we observe that all the trajectories collapse onto an equilibrium manifold before they end at the equilibrium position. The equilibrium manifold is obtained by connecting the different trajectories over the entire cross section and is shown by a dashed light-blue line in the figure. 
Here, we also display isocontours of constant total shear rate (dotted lines). Interestingly, these are tangential to the manifold in the vicinity of the equilibrium positions. This indicates that once the particle starts the secondary slow motion towards the equibrium, they move along lines of constant shear-rate. The isolines of total shear rate that are tangential to the manifold close to the longer and shorter wall have values of  $0.75$ and $0.81$ in units of $U_b/H$.

Next, we compare the features of the particle inertial migration in square and rectangular ducts. First, we note that if the particle is released close to the duct centre and focuses at the equilibrium position at the centre of a longer wall, it does not experience the initial lateral motion towards a shorter wall. This is directly connected to the weaker shear-induced lift force toward the shorter walls. The opposite is true if the particle starts close to the centre and focuses at the equilibrium position at the centre of a shorter wall, see the purple line in the figure, in which case the initial lateral motion toward the longer wall is significant. 
Secondly, we report that the focusing length is longer for particles in a rectangular duct, about three times longer than that in square duct if the particle starts at the same lateral distance from the equilibrium position {, i.e. see figure \ref{fig:fig3}b.}
{This is attributed to the lower average shear rate in the rectangular duct, $\overline{\dot\gamma}_{rec}=2U_b/D_h|_{rec}={3U_b}/2H$, with respect to that in the square duct, $\overline{\dot\gamma}_{squ}=2U_b/D_h|_{squ}={2U_b}/H$. This induces weaker lateral forces on a particle in the former case.}

\subsection{Migration dynamics of a oblate particle}

In this section, we vary the shape of the particle and investigate the inertial migration of a spheroid oblate particle. Shape is an important biomarker that can be employed in several biological and industrial applications such as cell separation, diagnosis and food industry \cite[][]{Masaeli12}. 
Only few studies have been conducted so far concerning the equilibrium position of non-spherical particles in microfluidic configurations. 
In the experiments by \cite{Hur11} it is reported that the rotational diameter, $D_{max}$, of the particle, determines its final equilibrium position in a rectangular duct. Later on \cite{Masaeli12} have demonstrated that the equilibrium position of a rod-like particle is closer to the centre than that of a spherical particle and eventually the rod exhibits a tumbling motion around its shorter axis of symmetry. The spherical shape of the particle has been exploited in previous studies to develop theories for the force prediction so that the inertial migration of the non-spherical particles and the underlying relevant physical mechanisms are still unexplored. We thus employ numerical simulations to study the motion of an oblate particle in straight ducts. Note that here an additional degree of freedom, the particle orientation, has to be included in the analyses.

\subsubsection{Oblate particles in square ducts}

\begin{figure}
\begin{center}
   \includegraphics[width=12cm]{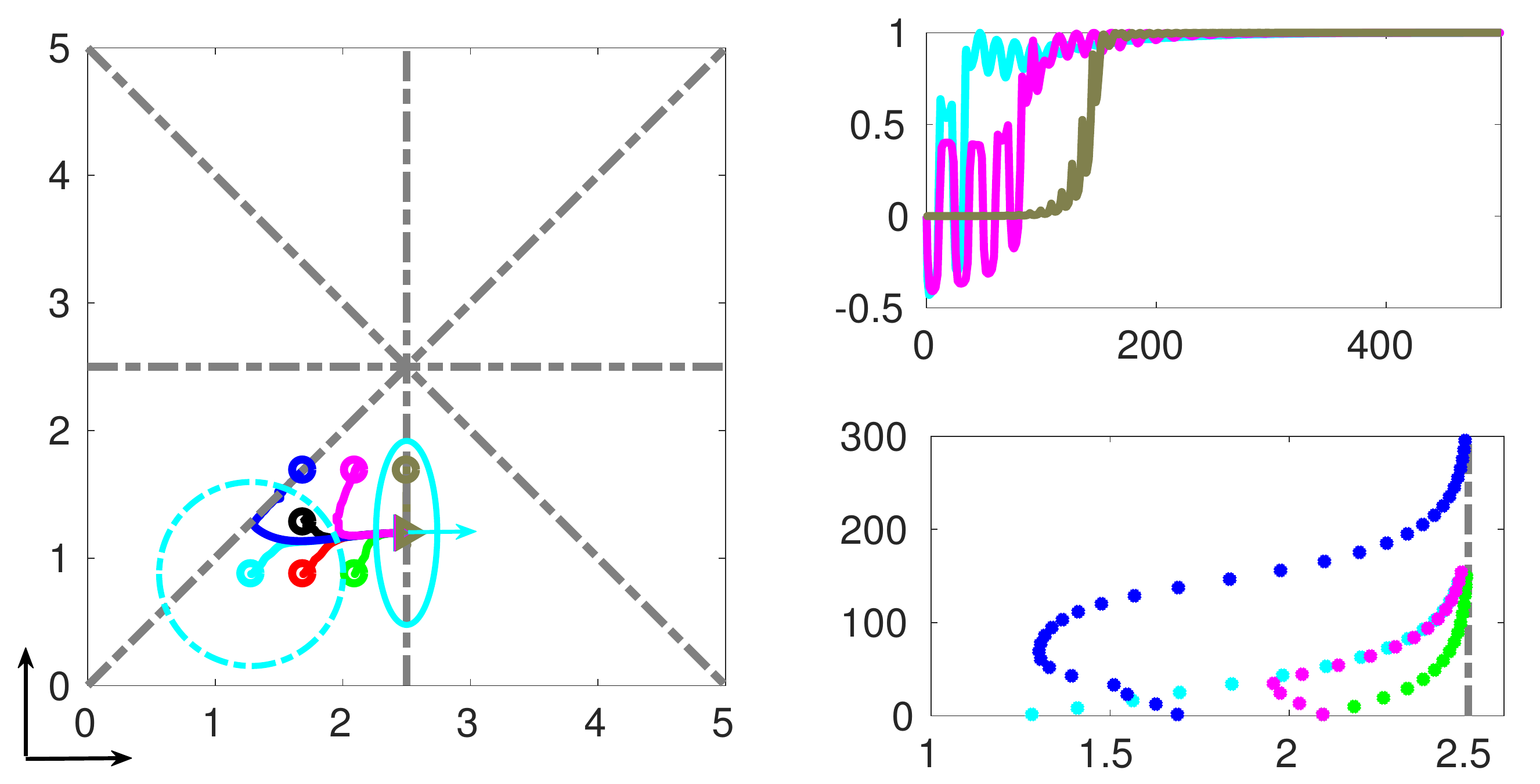}
   \put(-330,173){{\large $(a)$}} 
       \put(-305,5){{\large $x$}} 
        \put(-340,35){{\large $z$}} 
        \put(-235,55){{\large $\textbf{n}$}}       
    \put(-155,173){{\large $(b)$}}     
         \put(-5,45){{\large $y/H$}}  
         \put(-5,130){{\large $n_x$}}   
    \put(-152,82){{\large $(c)$}} 
         \put(-80,0){{\large $x$}} 
         \put(-70,90){{\large $t$}}
   \caption{a) Lateral trajectories of a rigid oblate particle over $1/8^{th}$ of a square duct at $Re_b=100$. The effective particle diameter is $1/5$ of the duct height(width), see definition in the text. The open circles and triangles show the initial position and final  equilibria, respectively. The initial and final orientation of the oblate are shown for a particular case with the light-blue solid line.{ b) Orientation of the oblate symmetry axis as a function of time scaled by $D^{e}_o/U_b$. c) Focusing length of oblate spheroids from the different simulations: the symbols represent the particle centre position at equal intervals of time.}}
   \label{fig:fig4}
\end{center}
\end{figure}
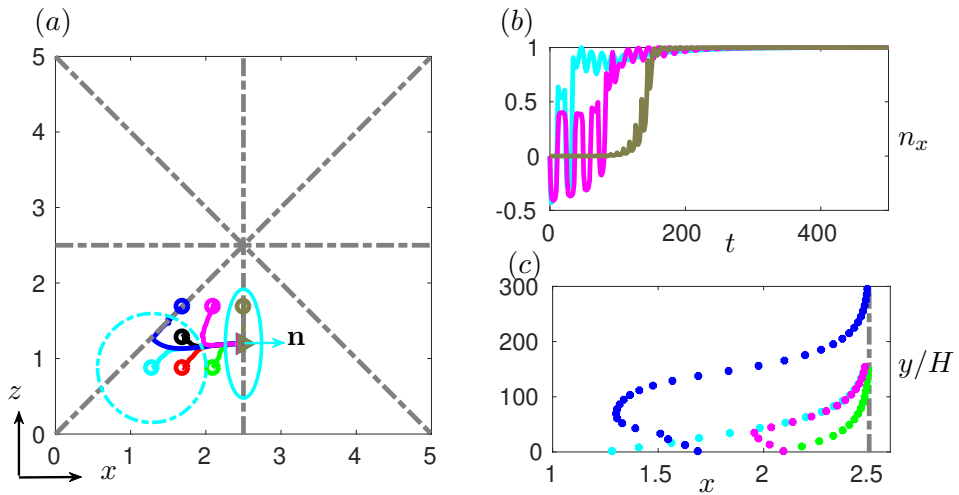

As for spherical particles, we first simulate the motion of the oblate spheroid in a square duct. The majority of the simulations have been performed for an aspect ratio of the oblate, polar over equatorial radius,  $AR=1/3$ (In {Appendix B} we report the effect of aspect ratio on the particle inertial migration). In addition we set the ratio between the duct height and the effective diameter of the oblate, the diameter of a sphere with the same volume, equal to $H/D^{e}_o=5$. 
This corresponds to a ratio between the duct height and the diameter of the oblate, the largest dimension of the oblate, equal to $H/D_o \approx 3.466$. We keep $Re_b=100$, $\overline{Re_p}=\overline{\dot \gamma} (D^{e}_o)^2/\nu=8$,  as for the simulations of a sphere discussed above and vary the initial position of the oblate centre over $1/8^{th}$ of the domain. The orientation of the oblate is monitored by the unit vector $\textbf{n}=[n_x,n_y,n_z]$ parallel to the oblate symmetry axis. Similarly, the particle rotation vector is defined by $\omega=[\omega_x, \omega_y, \omega_z]$. 
The particle initial orientation is $\textbf{n}=[0,1,0]$ for all simulations, indicating that its axis of symmetry is parallel to the streamwise direction whereas the initial velocity and rotation of the particle are zero.

By symmetry, we show in figure \ref{fig:fig4}(a) the lateral trajectory of the oblate only over $1/8^{th}$ of the duct cross section. The initial orientation and the actual size of the oblate are depicted by the light-blue dashed circle. 
The results of the simulations reveal features similar to those observed for spherical particles. 
In particular, 
the oblate particle experiences a lateral motion towards a face-centred equilibrium position so that four face-centered equilibrium positions are obtained as for spheres.
We also observe an initial lateral motion of the oblates towards a side wall, if they are positioned close to the centre, and  the collapse of the trajectories on the equilibrium manifold before reaching the equilibrium position. At the final equilibrium position, the orientation vector is $n=[1,0,0]$ for the face-centred equilibrium position shown in the figure regardless of the initial position. This is in agreement with the experimental finding by \cite{Dicarlo07} on inertial focusing of particles in microchannels and can be related to energy dissipation, as this is a configuration of lowest dissipation for oblate particles in shear flow. The oblate centre focuses about  $0.26H$ from the duct centre, slightly closer to the centre than a sphere at the same bulk Reynolds number.

The streamwise motion of an oblate spheroid, from the initial to the final equilibrium position, is accompanied by rotation around its main symmetry axis (polar axis) and tumbling around an equatorial axis. Initially, particle exhibits a Jeffery-like orbit \cite[][]{Jeffery22} with two distinct motions. First, a long-time motion without tumbling where the main symmetry axis is positioned perpendicular to the flow stream. Second, a tumbling motion in the form of a sudden rotation around an equatorial axis. As particle gets closer to the equilibrium position the tumbling motion vanishes gradually and particles just rotates around their axis of symmetry while travelling downstream. {In figure \ref{fig:fig4}(b)} we report the time evolution of the x-component of the orientation vector, $n_x$, for some of the simulated cases. {In this plot as well as the following plots, time is scaled by $D^{e}_o/U_b$. For a particular trajectory, indicated by the pink line, the first stage of the motion displays significant tumbling toward the equilibrium manifold,  $0< t < 110$, while the second has negligible tumbling along the manifold, $110< t < t_{final} \approx 500$}. 
The data in the figure indicate that longer time is needed to reach to the final orientation for particles starting close to the centre. Finally, we note that at the equilibrium position an oblate particle rotates around its symmetry axis. For the configuration in the figure,  $\omega=[-0.395,0,0]$ where the values are normalised by $U_b/D^{e}_o$.

The focusing length of oblate particles in square ducts are displayed {in figure \ref{fig:fig4}(c)}. Comparing to spherical particles of same size ratio, the focusing length is {slightly} longer for oblate particles when both are released from the same initial lateral position. This is mainly attributed to the tumbling motion in the cross flow direction which effectively reduces the lateral force on the oblate. {The symbols in the plot indicate the position of the particle center at equal time intervals showing the two-stage particle migration, the fast and slow motions, similar to the case of spherical particles.}

\subsubsection{Oblate particles in rectangular ducts}

\begin{figure}
\begin{center}
   \includegraphics[width=11cm]{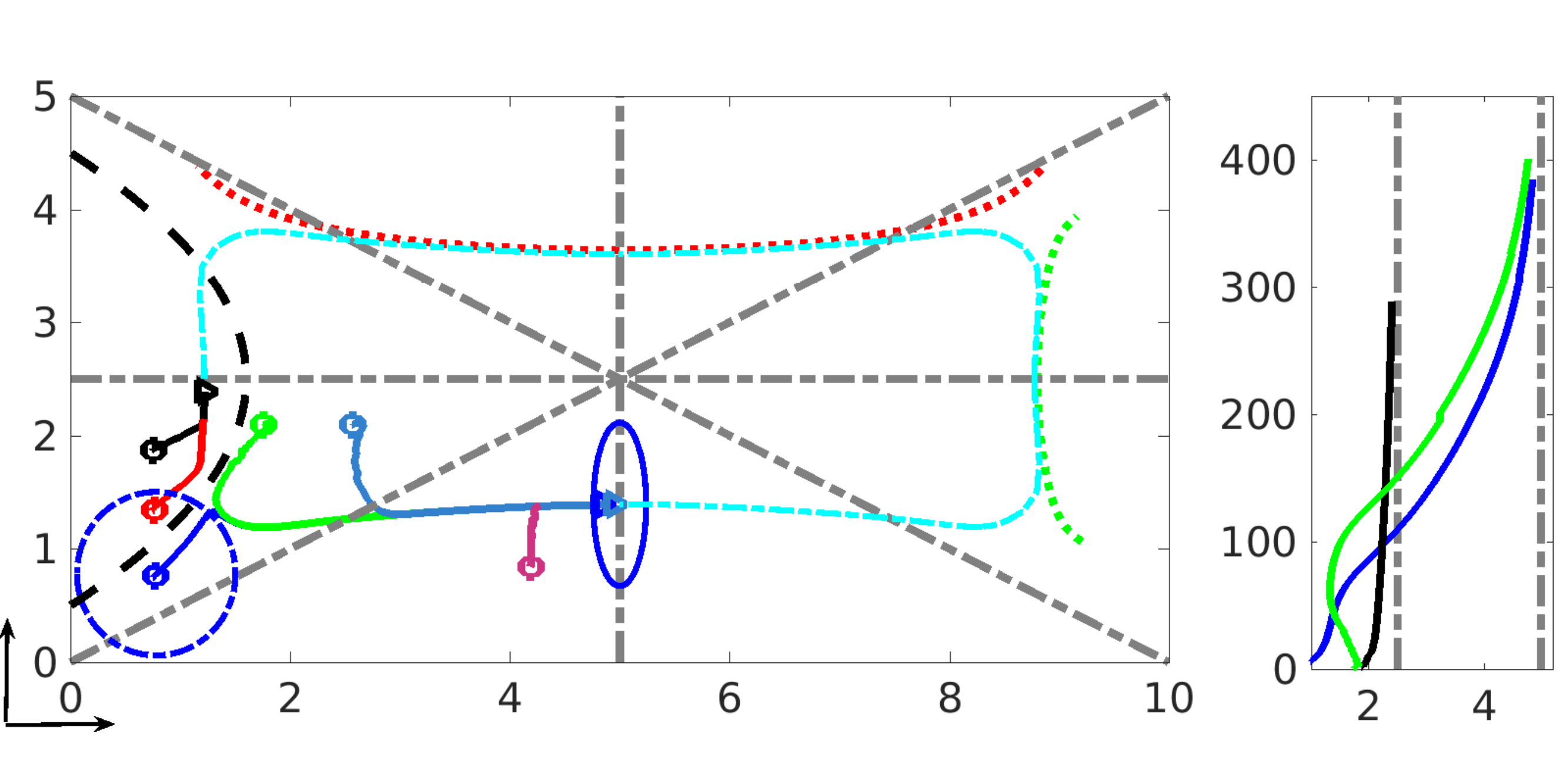}
       \put(-285,10){{\large $x$}} 
      \put(-315,40){{\large $z$}} 
     \put(-310,145){{\large $(a)$}} 
    \put(-60,145){{\large $(b)$}}  
      \put(-0,90){{\large $y/H$}}     
   \caption{a) Lateral trajectories of a rigid oblate particle transported in a rectangular duct, displayed over $1/4^{th}$ of the cross-section. The bulk Reynolds number  $Re_b=100$ and the effective particle diameter is $1/5$ of the duct height. Open circles and triangles show the initial positions and final equilibria from different realizations. The light-blue dashed line indicates the equilibrium manifold over the entire cross section, {that is where all the trajectories collapse before end at a equilibrium position}. The dotted red and green lines are  iso-levels of the total shear rate for the values of $0.67$ and $0.74$, chosen to be tangential to the equilibrium manifold (only shown on one side). The black-dashed  parabola defines two regions in the cross-section according to the final equilibrium position reached by particles released on each side of it. The two equilibria are at the face-centre of the long or short wall as indicated by the individual particle trajectories (only shown in one side). {b) Focusing length for different particle trajectories. The dashed lines represent the horizontal (z = 2.5) and vertical (x = 5) duct symmetry line, as the particle focuses on either of them.}}
   \label{fig:fig5}
\end{center}
\end{figure}
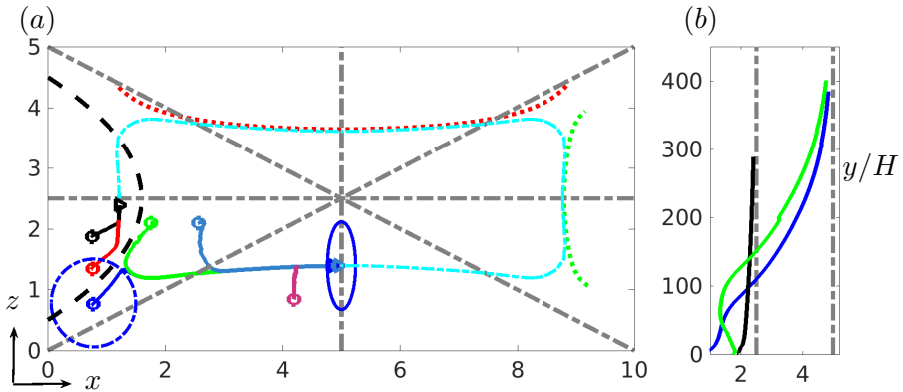

Next, we discuss the motion of an oblate spheroid in a rectangular duct of aspect ratio 2 as for the simulations of spherical particles discussed above. We also use the same  particle size ratio and bulk Reynolds number, $H/D_e=5$ and $Re_b=100$. 
The lateral motion of the oblate is displayed in figure \ref{fig:fig5}(a) over $1/4^{th}$ of the cross section. As in the case of a square duct, the particle reaches a steady configuration on the duct symmetry line at the center of the different faces. In addition, the particle symmetry axis, which is initially set parallel to the flow stream, $n=[0,1,0] $, finds its final orientation as $n=[1,0,0]$ for all the cases {if trajectories end at the face-centre of the long wall}, see figure. The distance of the oblate centre from the duct centre is $0.22H$ and $0.78H$ if it focuses at the equilibrium at centre of the long and short wall. The rotation vector at  the two equilibrium positions shown in the figure are  $\omega=[-0.34,0,0]$ and $\omega=[0,0.31,0]$ for the coordinate system adopted here.

Comparing these observations with the trajectories of a sphere in the same rectangular duct, we conclude that it is more likely for the oblate particle to focus close to the long wall, i.e. a larger portion of the cross-section will take the particle to the symmetry line orthogonal to the duct long wall, see black dashed line in the figure. 
In other words, the region of the cross-section delimiting initial positions corresponding to the equilibrium position at the centre of the short sides  is significantly smaller for oblate particles than for spheres. 
{We also} note that, in analogy to the case of spherical particles,  the equilibrium manifolds along which all trajectories collapse before reaching the final state are tangential to iso-lines of the total shear rate for the values $0.67$ and $0.74$. These  are slightly lower than the corresponding values for spherical particles as the oblate spheroids focus closer to the duct centre.   
{Finally, we show in figure \ref{fig:fig5}(b) the focusing length corresponding to some of the particle trajectories. The focusing length of the oblate in rectangular duct is shorter than that of a sphere in the same configuration. This is opposite to what we have reported for the particle motion in a square duct and is attributed to the higher lateral velocity of the oblate in the second stage of the migration, that with negligible tumbling.} 

\subsubsection{Size effect}

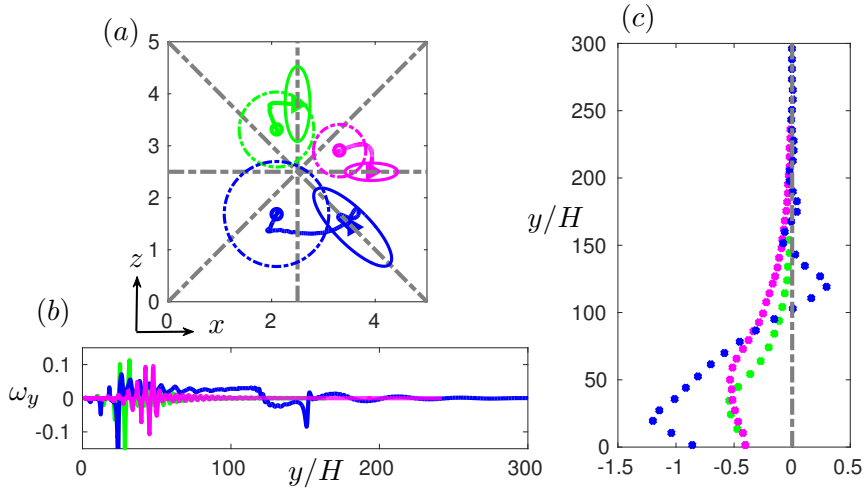
\begin{figure}
\begin{center}
   \includegraphics[width=12cm]{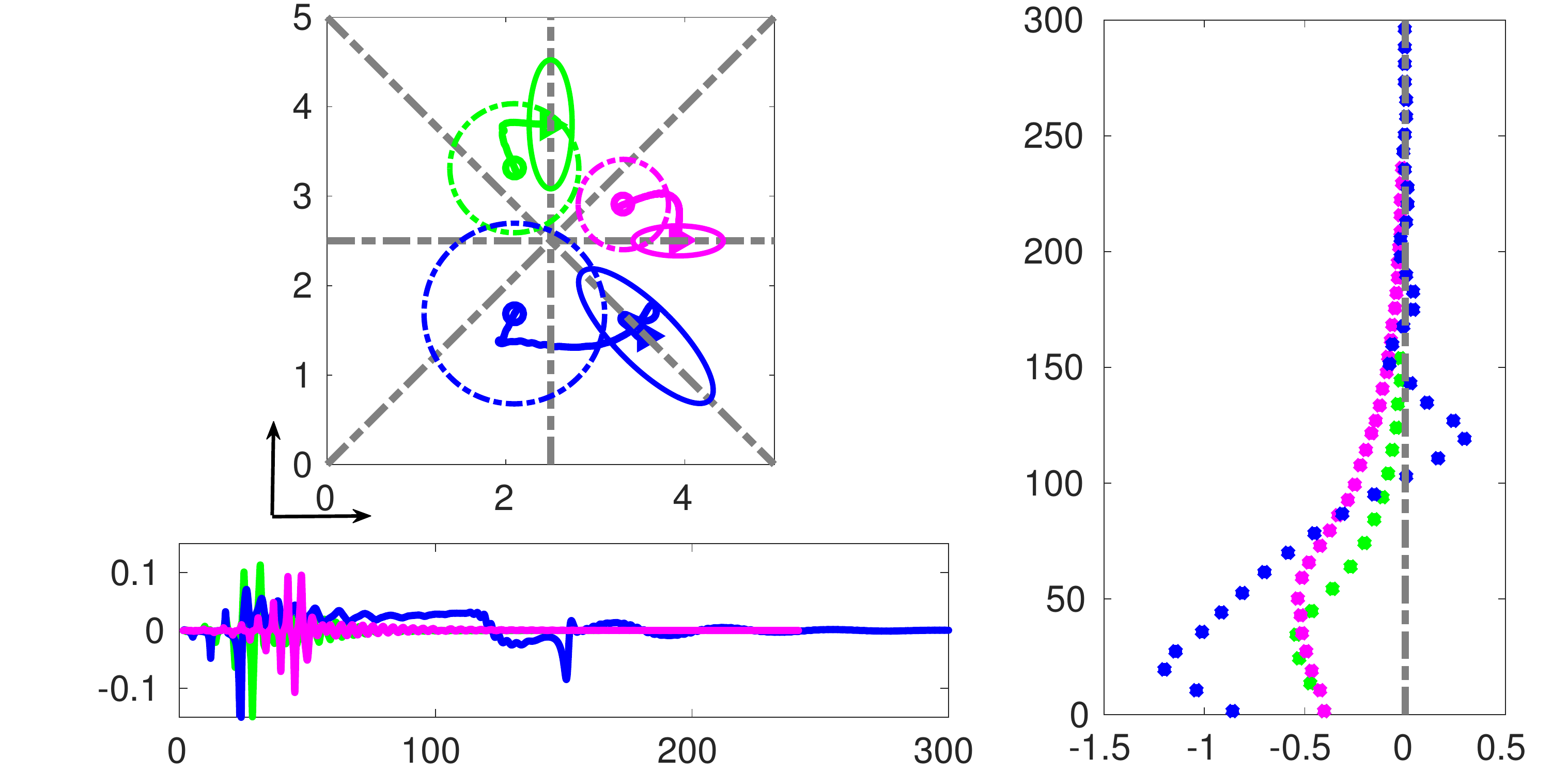}
   \put(-295,165){{\large $(a)$}} 
       \put(-255,55){{\large $x$}} 
        \put(-285,80){{\large $z$}} 
    \put(-320,60){{\large $(b)$}} 
     \put(-330,30){{\large $\omega_y$}} 
     \put(-225,0){{\large $y/H$}}   
        \put(-100,170){{\large $(c)$}}   
         \put(-135,95){{\large $y/H$}}
   \caption{a) Lateral trajectories of oblate rigid particles of different sizes in square duct at $Re_b=100$. The effective particle diameters are $1/10$, $1/5$ and $1/3.5$ of the duct height(width); Open circles and triangles show the initial and final equilibrium position respectively. The initial and final orientation of the oblates are shown with dashed and solid lines. b) Streamwise rotation-rate versus streamwise particle displacements for the three particles in (a), indicated by lines of the same color. c) Streamwise particle displacements  versus the vertical distance to the target symmetry line. }
   \label{fig:fig6} 
\end{center}
\end{figure}

As discussed in section \ref{SSC} particle size is an important parameter used in many biological applications. Here we therefore investigate 
how the particle size affects the particle trajectory and the ensuing equilibrium position for an oblate particle  in a square duct. 
Three different size ratios, $H/D_e=[3.5,5,10]$ are considered while keeping the bulk Reynolds number and oblate aspect ratio constant to $Re_b=100$ and $AR=1/3$. 
The lateral migration of the oblate spheroids in the square duct is depicted in figure \ref{fig:fig6}(a). Note that for each particle size, we performed several simulations, differing for the initial position over $1/8^{th}$ of the cross-section. In the figure, however, we only report the trajectories pertaining particles of different size released from the same position  (shown in different quadrants  of the cross-section for the sake of clarity). The initial and final configuration of the oblates are shown with dashed and solid lines respectively.  

First, we observe that the smallest oblate under consideration, $H/D_e=10$, exhibits a  behaviour similar to that of the reference oblate, $H/D_e=5$, discussed in the previous section. Regardless of the starting point, the oblate particle focuses at one of the four face-centred equilibrium positions with the symmetry axis, $\textbf{n}$, perpendicular to the duct symmetry line as shown in the figure. Interestingly, the behaviour is completely different for the largest oblate studied, $H/D_e=3.5$. Indeed, the equilibrium position is along the diagonal symmetry line, between the corner and the centre of the duct. In addition, we note that the oblate particle does not focus on the  diagonal closest to its initial position but on that on the opposite side. This  behaviour is consistently observed, for any initial condition considered over the cross section of the channel. 

  In order to explain the peculiar behaviour of the large oblate in square duct, we examine the rotation rate of the oblate with respect to the streamwise direction during the migration process, see figure~\ref{fig:fig6}(b). Typically, the rotation rate during the migration is initially  increasing and then decreasing. The former is associated to the fast motion of the particle towards the equilibrium wall and the latter to the particle slow motion towards the equilibrium position under the action of the rotation-induced lift force. This rotation-induced lift force reduces when decreasing the rotation rate and the particle eventually focuses at a face-centred equilibrium position. This behaviour is however not observed for the largest oblate under investigation. {In this case, the streamwise rotation does not reduce to zero as the particle moves towards the vertical symmetry line, at about $y/H \approx 100$ in the figure, which accelerates the particle motion. This behaviour might occur because the large oblate is more susceptible to tumbling,  something that we do not observe for spherical particles.} As a results of this acceleration, the particle crosses the vertical symmetry line and focuses on the diagonal where the streamwise rotation rate becomes zero and the shear-induced and wall-induced lift forces balance each other. 

To conclude the analysis, we report the focusing length of oblate particles of different sizes in \ref{fig:fig6}(c). Note that here we display the streamwise length as a function of the vertical distance to the target symmetry line for each case (vertical symmetry line for $H/D^{e}_o=[5,7]$ and diagonal symmetry line for $H/D^{e}_o=3.5$). It is evident from the figure that the smaller particles need longer times, and therefore travel longer in the streamwise direction, to reach their final equilibrium, similarly to what observed for spherical particles. It is also interesting to note that the centre of the large oblate cuts the target symmetry line several times while focusing, a behaviour always observed, regardless of the initial position of the particle. To summarise, we report in table \ref{tab:tab2} the final equilibrium position, orientation and rotation rate of the oblate particles in square duct at $Re_b=100$ as a function of the size ratio, $H/D^{e}_o$. 
We note that even though the smallest oblate, of size ratio $H/D^{e}_o=7$, focuses closer to the wall, where the magnitude of the background shear is higher, than {that of} the larger oblate ($H/D^{e}_o=5$), its final rotation rate is lower due to the hydrodynamic interactions with the wall.

 \begin{table*}
  \begin{center}
    \setlength\tabcolsep{1.5ex}
  \begin{tabular*}{\textwidth}{*{5}{c}}
    \hline
    $H/D^{e}_o$   & $\overline{Re_p}$ &     Distance from centre: (x,z)   &     Orientation: $\textbf{n}$    &      Rotation rate: $\omega$  \\ \hline
             3.5            &   16.33           &             (0.21H, 0.21H)               &  $[-0.71,0,-0.71]$        &         $[-0.33,0,-0.33]$          \\ \hline  
              5               &      8              &               (0, 0.26H)                 &        $[1,0,0]$                  &           $[-0.4,0,0]$           \\ \hline 
              7               &     4.08         &              (0, 0.29H)                &         $[1,0,0]$                  &         $[-0.32,0,0]$           \\ \hline   
  \end{tabular*}
  \caption{Final equilibrium position, orientation and rotation of an oblate rigid particle at $Re_b=100$ as a function of the size ratio, $H/D^{e}_o$}
  \label{tab:tab2}
\end{center}
\end{table*}

\subsubsection{Reynolds number effect}

\begin{figure}
\begin{center}
   \includegraphics[width=12cm]{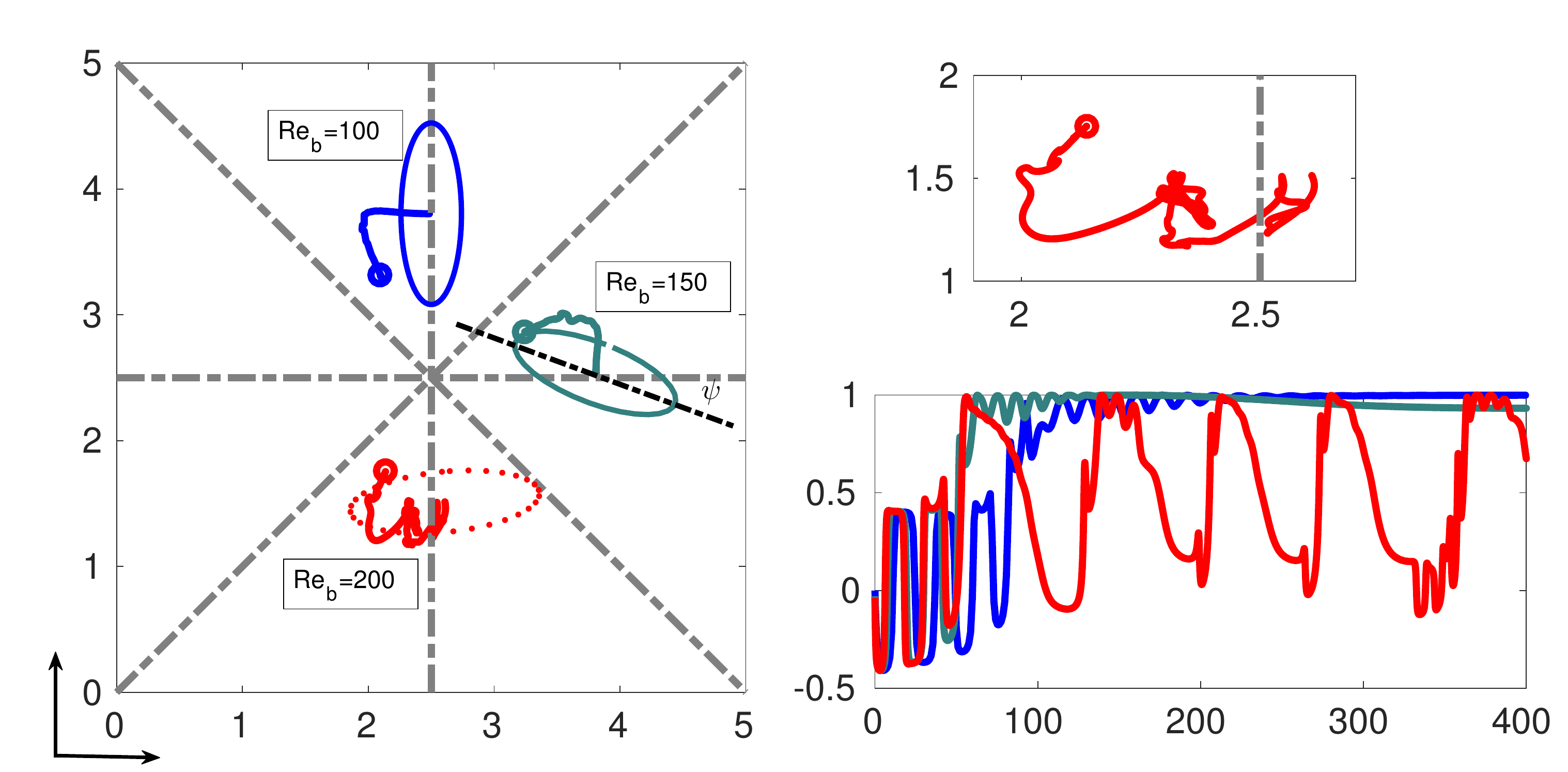}
   \put(-320,160){{\large $(a)$}} 
       \put(-305,0){{\large $x$}} 
        \put(-330,30){{\large $z$}} 
    \put(-160,160){{\large $(b)$}}      
      \put(-160,90){{\large $(c)$}}  
         \put(-5,50){{\large $n_x$}}
           \put(-95,0){{\large $t$}}
   \caption{a) Lateral trajectories of a rigid oblate particle, size ratio $H/D^{e}_o=5$, in a square duct at different bulk Reynolds numbers. Open circles and triangles show the initial and final equilibrium position, respectively. The final orientation of the oblate spheroid is shown with solid and dashed (for unstable cases) lines. b) Zoom of a particle trajectory at $Re_b=200$. c) The x-component of the orientation vector, $n_x$, as a function of time for the different Reynolds numbers under investigations (same color coding as in panel a). }
   \label{fig:fig7}
\end{center}
\end{figure}
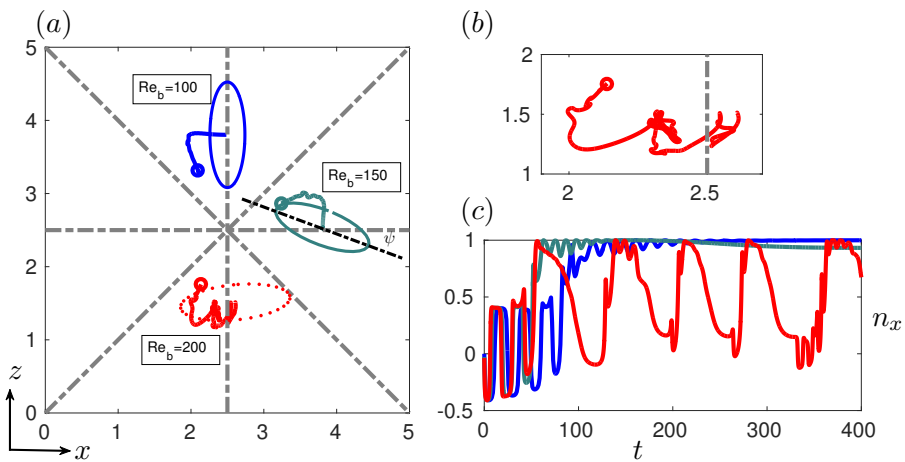

In this section we study the effect of the Reynolds number on the migration of an oblate particle in a square duct. We first note that we have run simulations at $Re_b<100$ and observed the same behaviour as at $Re_b=100$, with the only difference that the final equilibrium position is closer to the centre as observed for spherical particles, see as an example the data pertaining $Re_b=50$ in table \ref{tab:tab3}. 
Thus we will focus here on the results obtained at $Re_b=[100, 150, 200]$, see figure \ref{fig:fig7}. Note that we have performed more than two simulations for each Reynolds number changing the particle initial position, but we report in the figure only the trajectories originating from the same initial position and pertaining the different Reynolds numbers, displaying them over different quadrants of the cross-section for the sake of clarity.

At $Re_b=150$, the oblate moves laterally toward a face-centred equilibrium position and eventually focuses near (not exactly at) the duct symmetry line. In addition, at the final configuration, the particle remains tilted, at an angle $\psi \approx 21^{\circ}$ with respect to the duct symmetry line, see \ref{fig:fig7}(a). This configuration is independent of the initial position of the oblate and mimics the inclined rolling dynamic of an oblate in a simple shear flow, see table \ref{tab:tab3} for the particle specification.
{It is described by a pitchfork bifurcation in the dynamical system analysis of the motion of an oblate spheroid in a simple shear flow \cite[see for more details ][]{Ding00,Rosen15}. We note that the same behaviour is observed when we increase the streamwise box length from $6h$ to $8h$ and therefore we believe that  this effect is not related to the periodicity of the domain in the streamwise direction.}
{Finally, we report that the focusing length of the oblate spheroid at $Re_b=100$ and $Re_b=150$, for the trajectories shown in figure \ref{fig:fig7}, are $190H$ and $140H$, respectively. Therefore, the reduction in the focusing length by increasing the Reynolds number is also evident for the oblate particles.}

At $Re_b=200$ however the particle dynamics changes again. 
The oblate particle exhibits time-dependent rolling and tumbling motions around an equilibrium position at the face centre.
Therefore, we cannot define a steady orientation and rotation-rate at $Re_b=200$. 
We display one of  the orientations of the oblate in the lower part of figure \ref{fig:fig7}(a) using dotted line. The zoom on the particle trajectory is shown in figure \ref{fig:fig7}(b) where its chaotic dynamic is evident. A similar behaviour is obtained for particles transported at $Re_b> 200$,  not shown here.  The occurrence of a combined rolling and tumbling motion of an oblate at particle Reynolds number above a critical threshold is reported among others in the recent work by \cite{Rosen15} who considered homogeneous shear flow. 
Here we observe the chaotic motion of the oblate for $\overline{Re_p} \approx 16$. This may appear in contradiction  with the results of the previous section for the large oblate, $H/D^{e}_o=3.5$ at $Re_b=100$, where the $\overline{Re_p}$ shares similar value. We believe that two following reasons could possibly prevent the large oblate from tumbling at its final configuration:  1) The geometrical confinement  when the large oblate focuses close to the corner and 2) The weaker background shear  (the background flow shear-rate is much lower at the corners) which induces a lower local (effective) particle Reynolds number than $\overline{Re_p} \approx 16$.

Finally we show in figure \ref{fig:fig7}(c) the time evolution of the x-component of the oblate symmetry vector, $n_x$,  for each Reynolds number under consideration. It is evident that at the final focusing  $n_x$ is constant but smaller that unity for $Re_b=150$, whereas it  does not converge to the a particular value when $Re_b=200$. For the specification of the oblate particle orientation and velocities at the final position we refer the readers to table \ref{tab:tab3}.

 \begin{table*}
  \begin{center}
    \setlength\tabcolsep{1.5ex}
  \begin{tabular*}{\textwidth}{*{6}{c}}
    \hline
    $Re_b$   & $\overline{Re_p}$    &    Distance from centre: (x,z)   &     Orientation: $\textbf{n}$    &      Rotation rate: $\omega$  \\ \hline
           50            &       4          &          (0, 0.257H)                        &                  $[1,0,0]$       &          $[-0.41,0,0]$          \\ \hline
           100            &       8          &     (0, 0.26H)                        &                  $[1,0,0]$       &          $[-0.4,0,0]$          \\ \hline  
           150            &      12        &     (0.01H, 0.261H)                        &   $[0.93, -0.06, 0.36]$                  &        $[-0.32,0.02,-0.13]$           \\ \hline 
            200            &     16         &    $(\approx 0, \approx 0.23H)$     &          Time-dependent                  &         Time-dependent           \\ \hline   
  \end{tabular*}
  \caption{Final equilibrium position, orientation and rotation of an oblate particle of size ratio $H/D^{e}_o=5$  transported in a square duct as a function of the bulk Reynolds number, $Re_b$}
  \label{tab:tab3}
\end{center}
\end{table*}

 
 \section{Conclusions and remarks}
We study numerically the inertial migration of a single rigid particle in straight square and rectangular ducts. An interface-resolved numerical algorithm, based on the Immersed Boundary Method, is employed to study, for the first time, the entire inertial migration of an oblate particle in both square and rectangular ducts and compare it with that of a single sphere. The influence of the particle initial position, the size ratio and the Reynolds number on both particle lateral trajectory and its final,  equilibrium, position are investigated and documented. The results shed some light on the physical mechanisms behind the particle lateral motion and may suggest novel criteria to design microfluidic devices for separation and focusing purposes.  
 
First, we consider the inertial focusing of a sphere in both square and rectangular duct. We show that particles finds their equilibrium position at one of the four wall-centered symmetry lines, independently of their initial position.
 In the rectangular duct, however, it is more likely for the sphere to focus at the centre of the longer wall.  After an initial faster migration, the particle lateral trajectories are found to collapse onto a single line, denoted as equilibrium manifold, thus reaching the equilibrium position. The manifold is found to be tangential to the iso-levels  of the flow total shear rate in both square and rectangular ducts. In addition, spheres of larger size focuses closer to the duct centre whereas changing the Reynolds number varies non-monotonically the distance between the equilibrium position and the duct centre.   

As regards the lateral motions of a single oblate particle in a square and rectangular duct, we observe a behaviour qualitatively similar to that of a sphere in terms of final equilibrium position and equilibrium manifold. {However, oblate particles experience tumbling motion throughout the migration process which induces non-uniform lateral velocities and longer downstream focusing length in the square duct.}
At the equilibrium, an oblate particle remains vertical with respect to the adjacent wall with zero streamwise and wall normal rotation rates. This behaviour is independent of the initial position and orientation of the oblate. In the rectangular duct, it is more likely for the oblate to focus at the face-centre of the longer wall, an effect also observed for spherical particles but more pronounced for oblates.
{Moreover, the higher lateral velocity of a oblate with respect to a sphere in the second stage of the migration results in a lower focusing length for an oblate in rectangular duct.}

We also study the effect of the particle size on the inertial migration of an oblate spheroid in a square duct. For the smallest oblate considered, of size ratio $H/D^{e}_o=7$, the dynamics is similar to that of the reference case, $H/D^{e}_o=5$, just discussed, the main difference being the longer focusing length needed by smaller oblates as observed for spheres. 
For the largest oblate particle examined, $H/D^{e}_o=3.5$, however, the migration dynamics are completely different. First, we observe that the particle finds its equilibrium on the duct diagonal with its main axis of symmetry perpendicular to the symmetry line. To reach this, the particle first moves laterally toward a face-centred equilibrium position and then accelerates due to tumbling around a streamwise oriented axis. This causes the oblate to pass over the duct symmetry line and focus on the diagonal where the wall and shear-induced lift force balance each other.  

Finally, we study the effect of Reynolds number on the inertial focusing of single oblate particles in a square duct. At $Re_b<100$, the behavior is  similar to that obtained for $Re_b=100$ discussed above. The dynamics is  however different at higher Reynolds number. At $Re_b=150$, the oblate particle still  focuses close to a face-centred equilibrium position but remains tilted with respect to the duct symmetry line. {This equilibrium position resembles the inclined rolling motion due to the pitchfork bifurcation as reported in the dynamical system analysis of the motion of an oblate spheroid in a simple shear flow \cite[see for more details][]{Ding00,Rosen15}}.
Further increasing the Reynolds number, at $Re_b=200$, we find that the particle approaches an equilibrium position at the duct symmetry line, but the orientation and rotation are time-dependent and chaotic; a combination of rotational and tumbling motions. This behavior reminds of the finding of \cite{Rosen15} where at particle Reynolds number above a certain threshold the tumbling motion  adds to the pure rotational dynamics of an oblate in simple shear flow. 

Finally we note that the present study is among the first to investigate the entire inertial migration of an rigid non-spherical particle in straight ducts employing a highly accurate numerical algorithm.
The numerical approach presented here can be extended to consider different and more complicated shapes as well as deformable particles \cite[][]{berthet2013single,Zhu14,pham2015deformation}. These efforts may lead to new and more efficient microfluidic systems aiming to sort and separate particles.

 \section*{Acknowledgement}
This work was supported by the European Research Council Grant no. ERC-2013-CoG-616186, TRITOS and by the Swedish Research Council Grant no. VR 2014-5001. The authors acknowledge computer time provided by SNIC (Swedish National Infrastructure for Computing) and the support from the COST Action MP1305: Flowing matter.
 
\begin{figure}
\begin{center}
   \includegraphics[width=12cm]{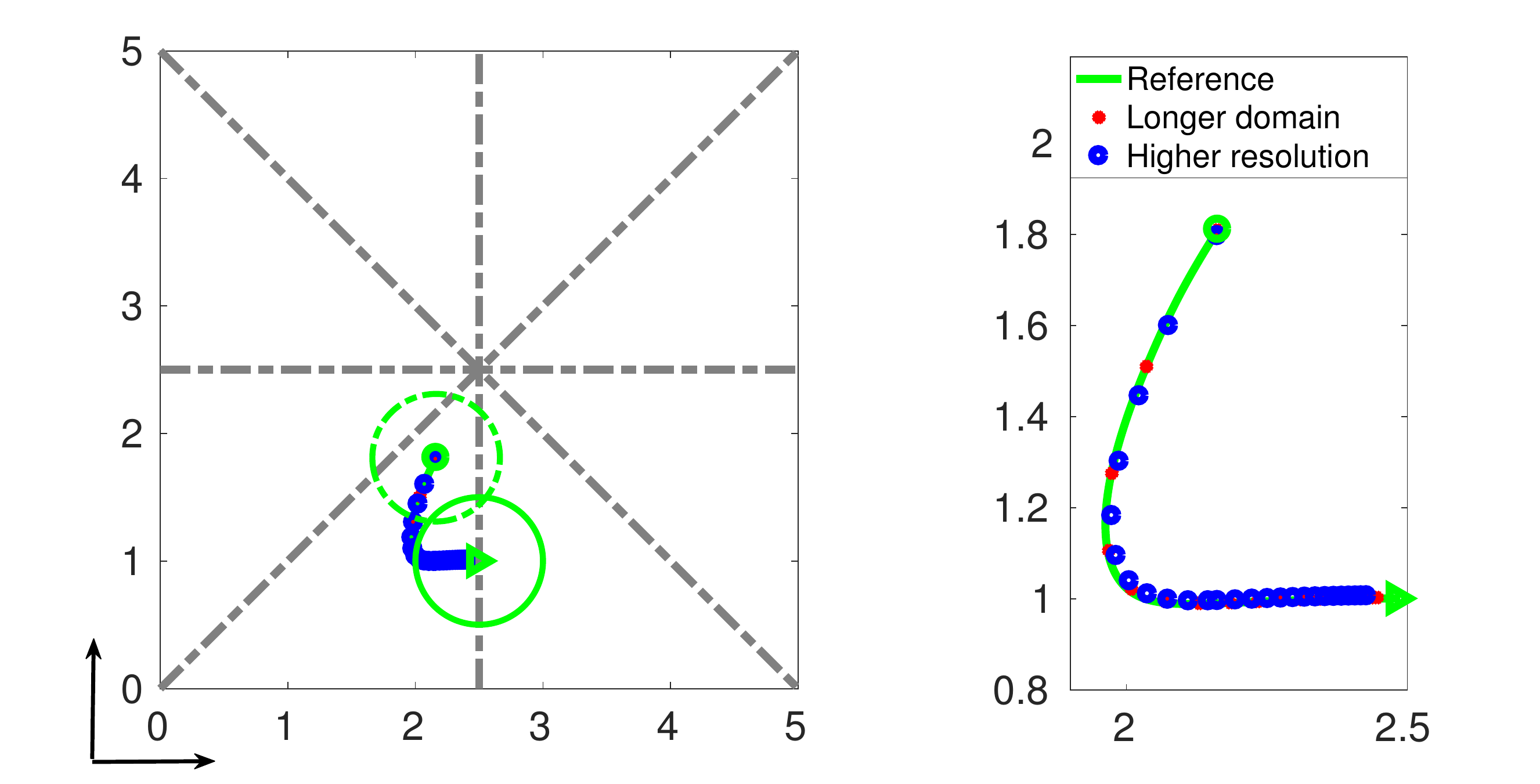}
    \put(-290,3){{\large $x$}} 
    \put(-325,35){{\large $z$}} 
     \put(-70,3){{\large $x$}}
     \put(-130,85){{\large $z$}} 
   \caption{ Lateral trajectory of a spherical particle with size ratio $H/D_s=5$ at $Re_b=100$ in the reference, longer domain and higher resolution flow cases}
   \label{fig:validation}
\end{center}
\end{figure}

 {
 \section*{Appendix A. Box size and resolution studies and validation} 
 In this part we examine the effects of box size and resolution on the particle dynamics. 
 In particular, we consider as reference case a square duct where the box size is $6h \times 2h \times 2h$ and corresponding number of grid points $480 \times 160 \times 160$ in the streamwise and cross flow directions. In the next step we increase the streamwise length from $6h$ to $8h$ to examine the possible effect of the streamwise periodicity on the particle migration dynamics. The case is denoted by "longer domain" where the grid spacing is kept the same as the reference case, i.e. $640 \times 160 \times 160$ grid points. Finally, we consider a third case where we keep the box size as the reference case and increase the resolution to $600 \times 200 \times 200$, denoted as "higher resolution". We examine the motion of a spherical particle, $H/D_s=5$, at $Re_b=100$ in the three cases mentioned above and report the results in figure \ref{fig:validation}. Note that in the latter case, 40 Eulerian grid points are considered per particle diameter and 5036 Lagrangian grid points cover the surface of the sphere. A very good agreement between the cases is obtained in terms of particle lateral trajectory and the final equilibrium position and therefore, we use the reference case throughout the study. \\
 Finally, we compare the equilibrium position of a spherical particle in a square duct with that reported in \cite{Dicarlo09}. Considering $Re_c=U_mH/\nu=80$ ($U_m$ is the maximum fluid velocity) and  particle size ratio $D_s/H=0.22$, the equilibrium position is reported at 0.29H from the duct centre in figure 1 of the reference which is in a perfect agreement with our simulation.  } \\
\\ 
 
 \section*{Appendix B. Effect of particle aspect ratio } 
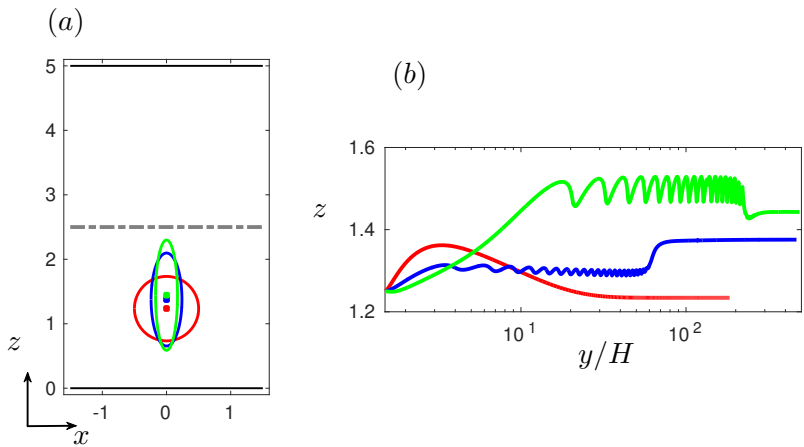
\begin{figure}
\begin{center}
   \includegraphics[width=11cm]{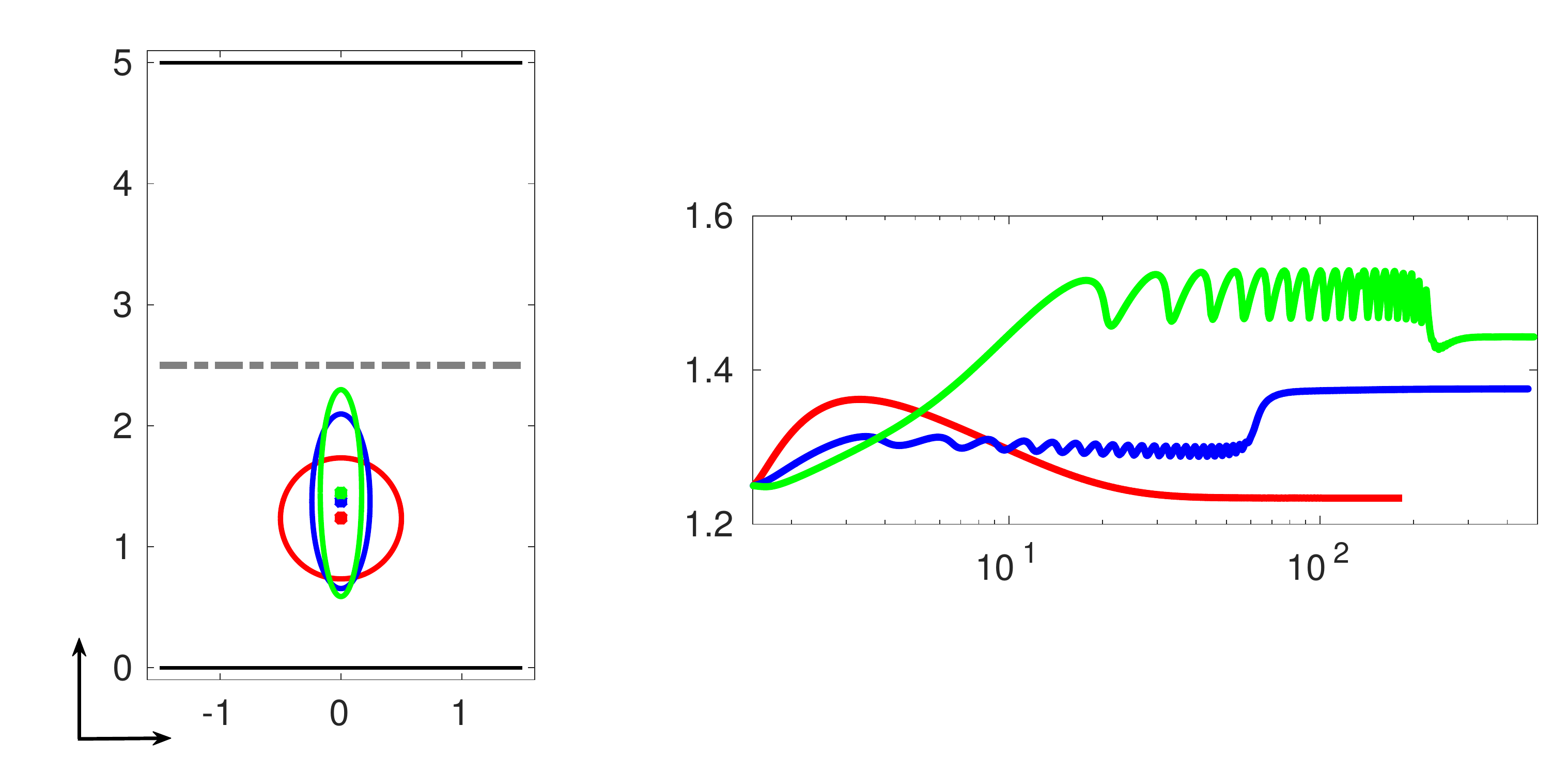}
   \put(-290,155){{\large $(a)$}} 
       \put(-280,0){{\large $x$}} 
        \put(-305,35){{\large $z$}} 
    \put(-160,135){{\large $(b)$}}     
         \put(-190,85){{\large $z$}}  
         \put(-90,30){{\large $y/H$}}
   \caption{a) Equilibrium position b) Focusing length of spherical and non-spherical rigid particles in a two-dimensional channel. The ratio between the (effective) diameter of the particles to channel height is equal to $1/5$ and the particle aspect ratio 1, 1/3 and 1/5. }
   \label{fig:fig8}
\end{center}
\end{figure}
In this Appendix we briefly discuss the effect of the particle aspect ratio on the equilibrium position and the trajectories of a single particle in a channel, a duct with infinite aspect ratio. In this case the flow is a pressure driven channel flow, i.e. a plane Poiseuille flow. The channel dimensions and the numerical resolution are chosen to be similar to those adopted for the rectangular duct. In this configuration, the particle migrates only in the wall normal direction, driven by the shear-induced and wall-induced lift forces \cite[][]{Matas04}.

We run simulations at  fixed $Re_b=100$ and $H/D^{e}_o=5$ and vary the particle aspect ratio in the different simulations. In figure~\ref{fig:fig8}, we show the trajectories, the equilibrium positions and the final orientation of spherical and oblate particles of aspect ratio $1/3$ and $1/5$. The sphere and oblate spheroids focus at a distance of $0.253H$, $0.225H$ and $0.211H$ from the centreline, respectively.  This result suggests that the more elongated the particles, the closer is their centre to the channel midplane at equilibrium and that the higher the duct aspect ratio, the closer the particles focus to the centreline. Note also that at the final equilibrium position, the oblate particles are oriented normal to the walls, as in a duct,  confirming that this configuration is not due to the presence of the side walls but related to the state of the lowest energy dissipation as mentioned above. We display the particle trajectories in figure \ref{fig:fig8}(b) where one can note that the oblate particles experience tumbling motion before reaching the final equilibrium positions. Similar to the flow cases in square duct, longer downstream length, focusing length, is needed for particles of higher aspect ratio. This is attributed to the tumbling motion which effectively reduce the particle lateral velocity.

\bibliography{mybibfile}
\bibliographystyle{jfm}

\end{document}